\documentclass[acmlarge, screen]{acmart}
\AtBeginDocument{%
  \providecommand\BibTeX{{%
    \normalfont B\kern-0.5em{\scshape i\kern-0.25em b}\kern-0.8em\TeX}}}

\setcopyright{acmcopyright}
\copyrightyear{2018}
\acmYear{2018}
\acmDOI{XXXXXXX.XXXXXXX}

\acmConference[Conference acronym 'XX]{Make sure to enter the correct
  conference title from your rights confirmation emai}{June 03--05,
  2018}{Woodstock, NY}
\acmPrice{15.00}
\acmISBN{978-1-4503-XXXX-X/18/06}

\usepackage{tabularx, multirow, graphics}
\begin{document}

\def \projectName {{AwareAuto}}

\title{Bridging the gap between natural user expression with complex automation programming in smart homes} 

\author{Yingtian Shi}
\authornote{Both authors contributed equally to this research.}
\email{shiyt0313@gmail.com}
\orcid{0000-0001-8733-7041}

\author{Xiaoyi Liu}
\authornotemark[1]
\email{liu-xy20@mails.tsinghua.edu.cn}
\orcid{0009-0001-3597-4783}
\affiliation{%
  \institution{Tsinghua University}
  \country{China}
}

\author{Chun Yu}
\authornote{Corresponding author.}
\orcid{0000-0003-2591-7993}
\affiliation{%
  \institution{Tsinghua University}
   \country{China}
}

\author{Tianao Yang}
\orcid{0009-0004-0593-3712}
\email{yta20@mails.tsinghua.edu.cn}
\affiliation{%
  \institution{Tsinghua University}
   \country{China}
}

\author{Cheng Gao}
\email{gaocheng20@mails.tsinghua.edu.cn}
\orcid{0009-0007-7583-7867}
\affiliation{%
  \institution{Tsinghua University}
   \country{China}
}

\author{Chen Liang}
\email{lliangchenc@163.com}
\orcid{0000-0003-0579-2716}
\affiliation{%
  \institution{Tsinghua University}
   \country{China}
}

\author{Yuanchun Shi}
\orcid{0000-0003-2273-6927}
\affiliation{
  \institution{Tsinghua University}
   \country{China}
}


\begin{abstract}
A long-standing challenge in end-user programming (EUP) is to trade off between natural user expression and the complexity of programming tasks. As large language models (LLMs) are empowered to handle semantic inference and natural language understanding, it remains under-explored how such capabilities can facilitate end-users to configure complex automation more naturally and easily. We propose \projectName{}, a EUP system in which the LLMs are introduced to the process of automation generation. \projectName{} allows contextual, multi-modality, and flexible user expression to configure complex automation tasks (e.g., dynamic parameters, multiple conditional branches, and temporal constraints) which are non-manageable in traditional EUP solutions. By studying complex rules data set, \projectName{} gains 91.7\% accuracy in matching user intentions and feasibility. \projectName{} effectively improves accuracy in reasoning and generating complex automation rules while facilitating natural user interaction. 
We discuss the opportunities and challenges of incorporating LLMs in end-user programming techniques and grounding  complex smart home contexts. 
 
\end{abstract}



\begin{teaserfigure}
  \includegraphics[width=\textwidth]{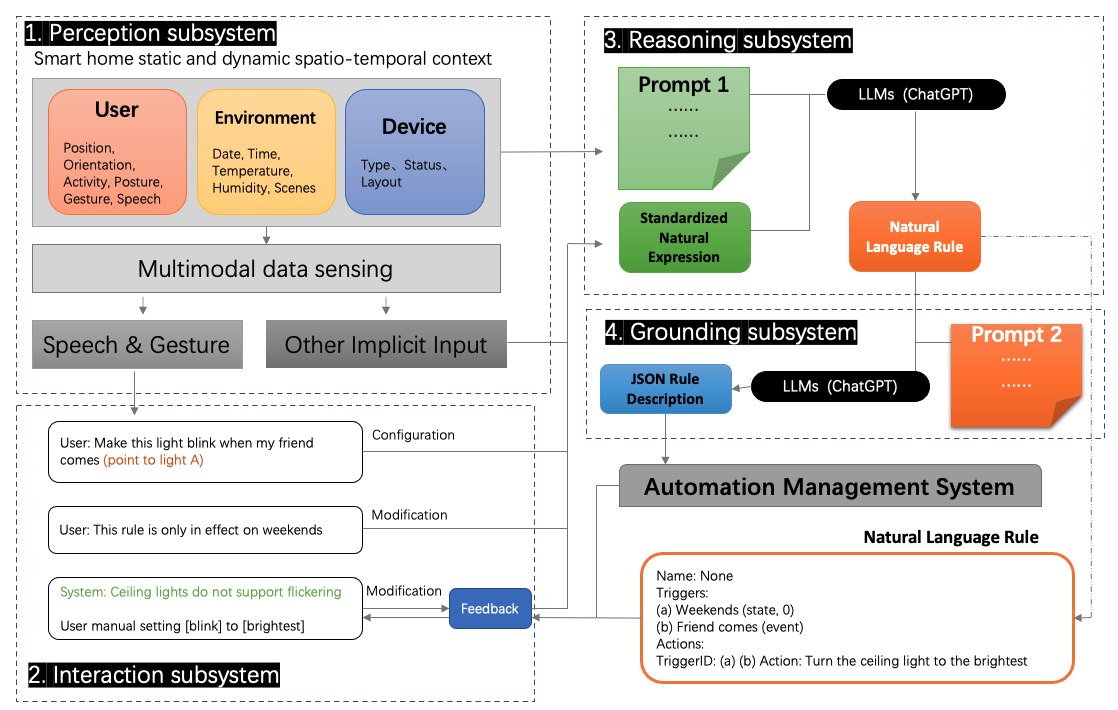}
  \caption{\projectName{} implements the reasoning process from natural user expressions to complex automation rules. The system first captures the multimodal input from the user and the context information of the smart environment. It completes a two-step inference for the user's intention and the feasibility of the rule with the help of LLMs. The reasoning result finally generates executable complex automation rules based on the interaction with the user.}
  \label{fig:teaser}
\end{teaserfigure}

\received{20 February 2007}
\received[revised]{12 March 2009}
\received[accepted]{5 June 2009}

\maketitle
\section{Introduction}
It is noted as a challenge to facilitate end users to program their smart home automation with natural and seamless interaction, especially with the increasingly complex IoT settings and social dynamics \cite{salovaara2021programmable}. 
End users tend to perform natural voice interaction and gestures in a flexible manner for rule expression, which creates ambiguity and uncertainty for the machine to map user intention with feasible automation programs. 
Although extensive literature explored end-user programming (EUP) methods to increase user experience or the expressiveness of complex rules, none combined the two as the ultimate goal. 
There is always a trade-off between the low-cost natural user expression and the complexity level of automation tasks. 
Either the proposed natural mechanisms are only applicable to trivial automation tasks, or the programming methods for complex tasks demand strict paradigm learning and constrain flexibility.


However, with the advancement of large language models (LLMs) such as GPT-3 \cite{brown2020language} in capacities of language learning and semantic reasoning, chances are that we can now bridge the gap between natural user expression and task complexity. 
It is definitely an exciting question of how, or to what extent, these capacities can augment a programmable smart home with human-AI collaboration. 
The challenges here are manifested in two aspects. 
\begin{itemize}
    \item \textbf{Naturalness Challenge}: To facilitate natural user interaction, \projectName{} provides a flexible, multi-model, and contextual programming paradigm \cite{ISP}. 
    This brings difficulties: 1) how to reason about user intents and the content of automation from ambiguous and information-limited user expressions \cite{liang2022code}, 
    furthermore 2) how to map the incomplete input into accurate output in terms of automation generation.
    
    \item \textbf{Grounding Challenge}: Smart home automation is a representative scenario for complex downstream reasoning tasks because the inference requires time-spatial, dynamic, and heterogeneous real-world context. 
    It also set barriers to LLMs feasibility due to hallucination \cite{welleck2019neural} and limited context length \cite{mialon2023augmented}.
    Hence, it remains to be addressed: 1) how to incorporate the complex inference information as input and 
    2) how to constrain the generated automation to be executable and feasible in real smart home context \cite{huang2022language}. 
\end{itemize}

To address the challenges, this paper proposed \projectName{}, an LLMs-based EUP method that maps flexible, multi-modal, and contextual user expressions to feasible automation programs grounding smart home context. 
Generally, \projectName{} integrates context-aware models, LLMs inference, and human feedback into the interactive programming process. 
The goal is to generate automation that is aligned with user intents and contextually feasible. 
We standardized the format of context-related user behavior and complex automation rules, then constructed a dynamic time-spatial prompt framework incorporating smart home context, real-time user interaction, and smart home affordances.
We developed \projectName{} system, which included contextual multi-modal sensing, reasoning, grounding, and interaction. 
We finally evaluated the generation performance and user interaction on the complex tasks of natural user programming interaction with 91.7\% 

Our contribution includes:
\begin{enumerate}
    \item We are the first to introduce LLMs technology in smart home end-user programming. We proposed \projectName{}, a EUP system for complex smart home configurations that combine LLMs inference with interactive programming process to facilitate user program complex smart home automation with natural expression. 
    
    
    \item We standardized the format for natural expressions and complex automation rules and designed a prompt framework to generate smart home automation. We defined the essential smart home contexts and representations for few-shots inference and aligned the LLMs inference to user intents and real-world feasibility.

    \item  We are the first to explore complex automation programming methods involving dynamic parameters, multi-modal input, and multiple conditional branches. We identified the challenges, developed methods, and gained  91.7\% inference accuracy in 9 types of complex automation generation. 

   \item We explored the LLM's capacities in grounding complex smart home 
   contexts and automation tasks. We investigated how to combine LLMs inference with context awareness and user interaction. We identified four errors of inference, including parameter splitting, inherent ambiguity, logical Correspondence, and illusion.


  
\end{enumerate}

Our work shows the opportunities to bridge natural user expression with complex automation programming. It is believed that \projectName{} opens up new directions for future EUP methods to balance naturalness and expressiveness.

\section{Backgrounds and Related Works}

\subsection{Naturalness and Expressiveness in End User Programming}
%

In the smart home field, end-user programming (EUP) aims to enable end users to tailor smart home automation \cite{barricelli2015designing}. 
The EUP tools are designed to facilitate users to configure the behaviors of smart devices through easy-to-use and easy-to-learn paradigms \cite{lieberman2006end,barricelli2019end} such as low-code visual programming languages \cite{ray2017survey} or programming by demonstration \cite{billard2008robot,chin2006end}. 
Trigger-Action programming (TAP) \cite{ur2014practical,ghiani2017personalization,huang2015supporting,ur2016trigger} and Event-Condition-Action (ECA) \cite{pane2001studying,paschke2006eca} paradigms are applied in the most popular commercial tools such as IFTTT \cite{ifttt}. 
Users can composite the automation rules in the form of ``if-triggers-then-actions'' or ``if conditions-when-events-then-actions''. 
However, with the growth of smart devices and interaction, automation programming is becoming more complex and interconnected \cite{lago2021managing,stein2016third}. 
The EUP tools are required to not only support simple and basic tasks but also facilitate users to build, control, and manage complex tasks and systems \cite{caivano2018supporting,jakobi2018evolving}. 
Advanced solutions like Node-RED \cite{nodered} are designed for complex IoT system behaviors, through which users composite by drag-and-drop nodes and workflows in a ``canvas'' interface. 
Kind of confusing what is this However used for, better make complex rules and nature expression as two key points.
However, the method demands non-technical end-users to understand the event logic and data flow \cite{Janssen2014VisualDM}. 
Several works highlighted the significance of balancing naturalness and simplicity to use with paradigm expressiveness for high-level complexity, semantics, and abstraction \cite{desolda2017empowering,jakobi2018evolving,huang2015supporting}. 
Some investigated the features of complex tasks and extended expressiveness of the rules. 
\citet{desolda2017empowering,salovaara2021programmable} outlined logical relations,time-spatial constraints, and social needs in rules expressiveness.  
\citet{seiger2015modelling} examined complex tasks which are high-dynamical, distributive, logical, and geographical IoT settings \cite{seiger2015modelling}.
\citet{huang2015supporting} investigated the ambiguous mental models of trigger types (states and events) and action types (instantaneous, extended, and sustained actions). 
\citet{ghiani2017personalization} highlighted that individuals have a personalized understanding of programming concepts or metaphors, indicating the significance of flexible and customizable paradigms. 
\citet{jakobi2018evolving,agadakos2018butterfly} investigated the causes and effects of actions to help users map events and operations with complex system behaviors. 
However, these methods sacrifice naturalness and flexibility in configuration and increased understanding load. 
Furthermore, extensive works explored natural language \cite{van2020convo,gordon2014steps,clark2016towards} in conversation-based and voice-based methods, such as InstructableCrowd \cite{huang2016instructablecrowd}, Heytap \cite{corno2020heytap} and Convo\cite{van2020convo}. 
These methods allow users composite IF-THEN rules by specifying their needs or rules contents through less constrained, semantic, and contextual expression \cite{highseman,ardito2020user,funk2018addressing,ariano2022smartphone}. \citet{ammari2019music} identified the lack of support for time-spatial contextualization and dynamic instructions in existing voice-based tools. CoMMA \cite{kim2022conversational} 
leveraged contextual information (position, time. and etc.,) for disambiguation strategies and iterative modification. Jarvis \cite{lago2021managing} is a conversational agent empowering non-trivial tasks with contextual contents (time, place), temporal constraints, and external events. Others explored multi-modal and situated interaction to enhance naturalness and simplicity in IoT-based configuration by combining natural language 
with gestures, activities, and physical operations \cite{kang2019minuet,ariano2022smartphone,lee2013tangible,chin2006end,dey2004cappella,li2018appinite,li2017programming}. \citet{ISP} investigated natural user behaviors for situated programming and highlighted users greatly leveraged situated context and flexible verbal specifications complemented by multi-modal interaction (e.g., selecting the devices by pointing). The author also found that users create rules with multi-modal, dynamic, and multiple conditional branches. Whereas, to the best of our knowledge, these types of rules are under-explored in existing multi-modal EUP methods. Moreover, for complex rules, the previous works mainly focus on simplifying rule specification \cite{desolda2017empowering} rather than natural and flexible mechanisms. 

In contrast, our research aims to improve the naturalness and simplicity of programming complex automation. For one thing, we aim to extend the expressiveness of the state-of-art multi-modal methods in complexity and diversity. We specifically explored three sources of rule complexity in rule parameters, correspondence between triggers and actions, and temporal logic. For another, we aim to improve naturalness and simplicity by allowing ambiguous, multi-modal, and flexible expressions by situated reasoning contextual and semantic information.

. 

\subsection{LLMs Grounding Real-World Tasks}
Large Language Models (LLMs) (e.g., BERT\cite{devlin2018bert}, GPT-3\cite{brown2020language}, 
LaMDA \cite{thoppilan2022lamda}), 
are trained on massive and broad data and can be fine-tuned to various downstream tasks, bringing great opportunities for human-computer interaction and more complex intelligent tasks \cite{bommasani2021opportunities,mialon2023augmented}, such as robot manipulation, reasoning, multi-model generation. Especially, GPT-3 \cite{brown2020language}, and the more recently released GPT-4 \cite{gpt4}, have achieved state-of-the-art performance to various ranges of real-world tasks without 
fine-tuning on a task dataset, but by only using in-context learning, also called prompts \cite{reynolds2021prompt} in the form of  natural language instructions. The unprecedented capacities of natural language understanding, semantic knowledge, and reasoning enable several exciting products such as ChatGPT \footnote{https://openai.com/blog/chatgpt}, New Bing \footnote{https://www.bing.com/new}.etc. However, the prompt techniques have limitations such as the limited tokens and sensitiveness of the input, various studies, therefore, investigated prompt design or prompt engineering \cite{betz2021thinking,qiao2022reasoning,liu2021makes,lu2021fantastically} to invoke desired functionality compatible with complex in-context tasks and active human-AI interaction. \citet{wei2022chain} proposed Chain-of thought (CoT), a few-shot prompting technique consisting of examples of the contextual tasks and intermediate reasoning steps from input to the desired output. 

Despite the unprecedented capabilities of LLMs, there are still huge challenges to grounding real-world applications \cite{mialon2023augmented}. For one thing, LLMs are suffered from hallucination \cite{welleck2019neural} issues, namely, the output is nonfactual or unrelated but seemingly plausible. Another hard nut is  
connecting LLMs with real-world environments, including digital environments (e.g., knowledge bases, websites. etc) physical environments \cite{gu2022don}, and social situations \cite{krishna2022socially}. Hence, an extensive body of literature explores methods to yield valid, factual, and contextually appropriate outputs. One direction is augmenting the output with external tools \cite{schick2023toolformer,shuster2022blenderbot} (e.g., external database, search engine, APIs, code interpreter, even human) or reinforcing learning with real-world affordance \cite{ahn2022can} or human feedback \cite{warnell2018deep}. Another direction is to leverage LLMs to act in the virtual and physical world, including virtual assistants, robots, and human interaction. \citet{li2022pre} encodes the tasks and virtual home environment with a templated English sentence. They also encoded the 3D information from the agent's observation and 3D world coordinates. \cite{liang2022code} proposed policy used prompting hierarchical
code-gen (recursively defining undefined functions), which allows spatial-geometric reasoning and ambiguous descriptions inference.
\citet{huang2022language} proposed methods to improve the executability and correctness of the actions. SayCan \cite{ahn2022can} used reinforcement learning (RL) methods to constrain the LLM outputs to be feasible and contextualize the physical environment. In this research, a robot can perform complex and extended temporal tasks based on the operations yield from the LLM likelihood and affordance function \cite{zeng2019learning}. PROGPROMPT \cite{singh2022progprompt} proposed a programmatic promoting structure for situated robot manipulation. The prompts include action primitives and available devices to constrain the output. LLM-Planner \cite{song2022llm} constructed a dynamic prompt related to the updating environment. Socratic Models \cite{zeng2022socratic} and NLMap\cite{chen2022open}, and Inner Monologue \cite{huang2022inner}
integrated a visual-language model (VLM) to incorporate contextual information, including the description of objects and scenes and human interaction. 

The before-mentioned works are mostly simulated and 
evaluated in a virtual or physical domestic environment, indicating the feasibility of LLMs grounding to the home context and time-spatial inference. Our research designed the reference methods based on the previous works in terms of grounding spatial context and feasibility. Further, we explored the LLMs capacities in end-user programming systems and extended the inference methods by incorporating real-time and situated user interaction.  

\subsection{AI-enabled Smart Home Automation Programming}

With the rapid advances of artificial intelligence (AI) and natural language processing (NLP), a significant body of work research started to investigate AI-infused and customized smart homes \cite{barricelli2022exploring}. Basically, there are two lines of research: 1) synthesizing or enhancing the automation generation, and 2) facilitating end-user programming (EUP) interaction. 

For the first line of research, previous studies on how to map natural language specifications and other multi-model inputs (routine trace, activities) to executable automation programs or IFTTT rules. Generally, there are two challenges. One is the  incomplete, ambiguous, and inaccurate human expression and the other is to constrain the output executable and contextually feasible in the smart home context \cite{quirk2015language,huang2022language}. \citet{quirk2015language} used a
semantic parsing method to map natural language TAP rules to 
 executable code. \cite{liu2016latent,yusuf2022accurate} used deep-learning architecture to generate TAP rules, while these methods have limited performance when the description is vague or complex. 
\cite{barricelli2022multi,zhang2020trace2tap} leveraged the multi-modal features (vision, speech, and device operations) offered by smart speakers to synthesize routines-related TAP rules. However, these works primarily use offline data rather than real-time information during smart home interaction. 
 
For the second line of the research, 
recent studies endeavor to improve user experience by incorporating AI in the EUP tools. Some \cite{corno2021users,corno2020heytap,lago2021managing} use conversation agents to allow configuration from user intentions and routine preferences. Basically, they build user profiles, device affordance, and intentions for semantic reference. But these systems are limited in complex rules and only apply language specification as input, lacking multi-modal information from the users.

We are the first to explore a EUP method to grounding LLMs in generating complex smart home automation. Our method differs in two aspects. First, we focus on the natural smart home interaction when configuring the rules. We incorporate multi-model and contextual information to reason user intents and contextual feasibility. 
Second, we considered complex automation tasks, derived from both flexible user expressions and complexity from diverse IoT-related rules.





 





\section{Identifying Challenges}
\label{challen}
Before delving into the system for automation generation from natural user expressions, it is essential to identify and address the challenges involved. This section will discuss the challenges related to user behavior understanding and complex rule reasoning and identify key issues based on interaction data analysis.

Previous research has provided valuable insights into how users express themselves when dealing with complex automated rules. To build on this knowledge, we utilized a dataset from a previous study \cite{ISP}.


The dataset is a text-based dataset of multi-modal user interaction in in-situ programming tasks where users program automation rules with natural expressions (voice, gesture, and tangible interfaces) in a physical smart home environment. The dataset consists of 216 tasks. The tasks range in complexity from the rule with a single trigger and action to rules with multiple logic, parameters, and dynamic features. Each task is annotated with natural user behaviors (interaction features, acts, intentions, etc.) and a user-desired TAP rule. Regarding our research goals, we only focused on the complex tasks in the dataset.
We recorded user expression behaviors and analyzed 205 instances of natural user behaviors to configure automation rules. We discussed the details of the challenges posed by natural user expressions and the complexity of automation rules below.

\begin{figure}[htbp]
\centering
\includegraphics[width=14cm]{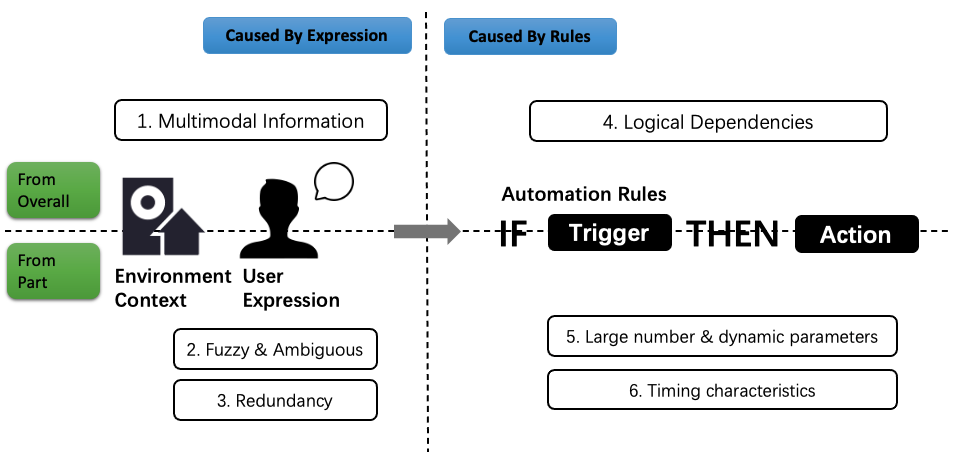}
\caption{Source of complexity for automation rule generation tasks. All sources can be classified into six categories, depending on where the complexity is generated and whether the complexity depends on holistic or partial factors.}
\end{figure}


%
\subsection{User natural behavior understanding}
The parameters and logic of the complex rules are all involved in the user's natural expressions, which can be ambiguous and incomplete. Hence, naturalness expressions create difficulties for parameter extraction and logic complementation. 
According to the behavior instances data, the sources of the challenges can be divided into three parts: multimodal expressions, ambiguous expressions, and redundant expressions.

The user's multimodal information includes active input, such as voice and gestures, and implicit context information, such as the user's activities and location. Active input is the primary source of logic and parameters for rule creation while implicit and environmental information supply auxiliary knowledge.

For instance, a rule expressed by the user is "(Point at ceiling light) Turn on this light at this time every day."
The user used multimodal and environmental information to simplify their expressions avoiding the complexity parameters such as the precise time and device name.
Identifying valuable users' multimodal information is the basis of parameter inference and is important for rule construction.

Besides multimodal and environmental information, there are also vague expressions, such as setting "turn up the air conditioning temperature" and "make the light warmer" without specific parameters, which poses a great challenge to the system's reasoning ability. The ambiguity includes the operation parameters and the device in the user's expression. Many users cannot say the exact name of the device to be operated but use some features to express the operation target, such as "sofa light". Such expressions may be ambiguous and need to be disambiguated by combining the contextual information of the situation and the user's expression. To address this challenge, we propose a solution in the subsequent design. Firstly, the system uses its reasoning ability to complete the parameters and then further clarifies the fuzzy parameters through interaction with the user.

There are missing and inaccurate parameters on users' multimodal and fuzzy expressions. Users' redundant expressions have unnecessary or supplementary information outside the rules. It is difficult for users to express the logic smoothly and clearly when they set up complex rules. They have modifications and additions in the expression process. Take the rule in the dataset as an example, the user's expression is "Set a sleep mode. If I lie on the couch, then all devices are off, and only the air conditioning is on. Nope, it doesn't need to be triggered by lying." where "user lies on the couch" is a redundant piece of information that may mislead the rule logic. The redundant information also poses a problem for the correct reasoning of the rule.

\subsection{Source of Complexity in Rules}
In addition to the challenges posed by user representation, the rules' complexity also poses a challenge to reasoning. Through statistics and previous literature\cite{lago2021managing,huang2015supporting,brich2017exploring}, we found that the complexity of rules is mainly focused on three aspects. 

The first is the complexity of the rule parameters. 
The complexity of an automation rule is highly related to the type and number of parameters and can be divided into three cases. For the first case, the more triggers in the rule, the more actions need to be executed, and the more complex the corresponding rule is. Rule complexity is mainly reflected in the number of parameters. For the second case, there are fuzzy parameters and spatially relevant parameters in the rules that need to be obtained by inference, and the appearance of this part is related to the natural expression of the user. In the third case, some complex rules have dynamic parameters. For example, one rule is "When the user sits down, turn on the light near the user." The target device in this rule is related to the user's location and will change dynamically with the scenario. 

The second is the correspondence between trigger and action \cite{lago2021managing,ur2014practical}. There may be different branching situations in a rule, such as ``Turn on the ceiling light when the user is watching TV and switch the light to warm if it is raining outside.'' Branches often exist in the natural expression of the user as a supplement to the rule or modification. The trigger of different branches can be complementary or recursive, and different triggers can correspond to the same action. A rule with branches can be split into multiple rules from the system's implementation level. Still, users usually tend to treat these rules as multiple branches of one rule when configuring them and treat them as one rule when managing and editing. Branching rules present a greater challenge at both the logical and system implementation levels.

The third is the temporal logic in the rules, such as "Turn off all lights after the user leaves the living room for ten minutes.", "Open the curtains before the floor robot starts working." Some triggers and actions in the rules have delayed time or strict execution orders. The temporal logic of the rules can be divided into two types: the trigger method and the execution order of the action. The past research divides the trigger into two categories: event and state. The event-type trigger is only triggered when the event occurs, such as "turn on the TV", while the state-type trigger requires a state rotation, and the rule is triggered when the state is satisfied. Some of the state rules need to be triggered after a period of time, such as "after the user leaves the living room for 10 minutes". Understanding the rule logic requires a correct judgment of the timing logic of the trigger to reason about the triggering method. The temporal complexity of Action is mainly reflected in the execution order, and the execution time of each operation needs to be reasoned correctly.

In many rule settings, the complexity due to different behaviors and characteristics will be combined together, thus increasing the complexity of the whole rule inference task. We have classified the data according to the complexity analysis described above. Of all complex rules, 28.3\% (58) are influenced by user expression and 62\% (127) by rule complexity. And 9.8\% (20) of these rules receive challenges from both cases. 

\section{Design and Implementations}
After analyzing the available data and identifying the challenges associated with complex rule generation, we developed a system that utilizes a large language model to facilitate natural user expression in the rule generation process. We have divided the system into four parts according to the automated rule generation steps and the system's core functions.
\begin{figure}[htbp]
\centering
\includegraphics[width=10cm]{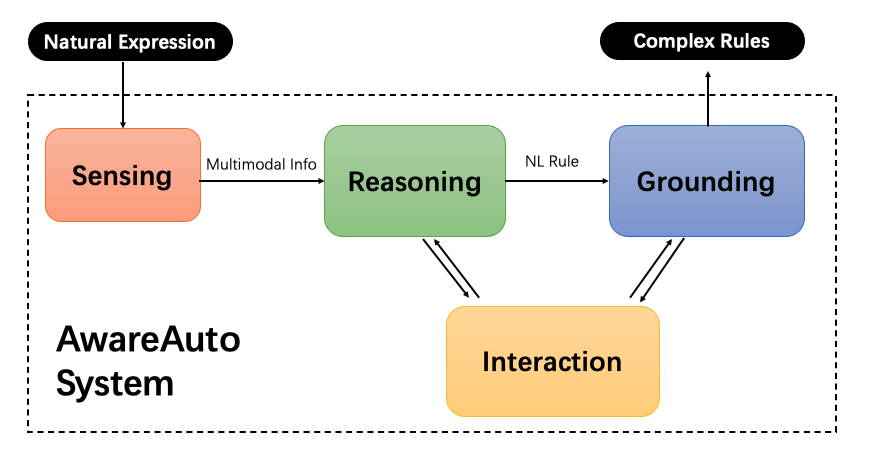}
\caption{\projectName{} System Framework. The cooperation between the four subsystems of Sensing, Reasoning, Grounding, and Interaction transforms natural user expressions into complex automation rules that can be executed.}
\end{figure}

The first part is the capture of the user's natural expressions through various sensors in the environment, such as verbal expressions, position, posture, gestures, and activities. This information is used to extract relevant data for rule generation. The standardization of user information provides the necessary parameter for subsequent reasoning.

The second part involves understanding the logic of the user's rules by synthesizing their natural expressions and environmental information. This part is achieved through prompt testing with the help of LLM, which helps to understand the parameters in the user's complex rules, such as the correspondence between the trigger and action and the order of execution of the action.

The third part involves deploying the rules by grounding the generated rule logic and the available interfaces of intelligent devices in the environment. With the help of LLM and user interaction, the system completes the final rule establishment.

The fourth aspect pertains to the interaction between the system and the user, which enables greater control over the process of automation generation. Through the interaction, users can rectify any misunderstandings or provide additional information to the system, resulting in the desired automation rules.

We will provide more detailed information about the design of each part of the system in the following paper.

\subsection{Standardization of multimodal information}

Capturing natural expressions from users is crucial for automation generation. However, due to the complexity and diversity of context information, acquiring relevant input information for constructing automation is essential. To achieve this, we have standardized user expressions based on data analysis. The standardized information comprises environment-related information, user posture and activities, and user-initiated expressions.
\begin{figure}[htbp]
\centering
\includegraphics[width=12cm]{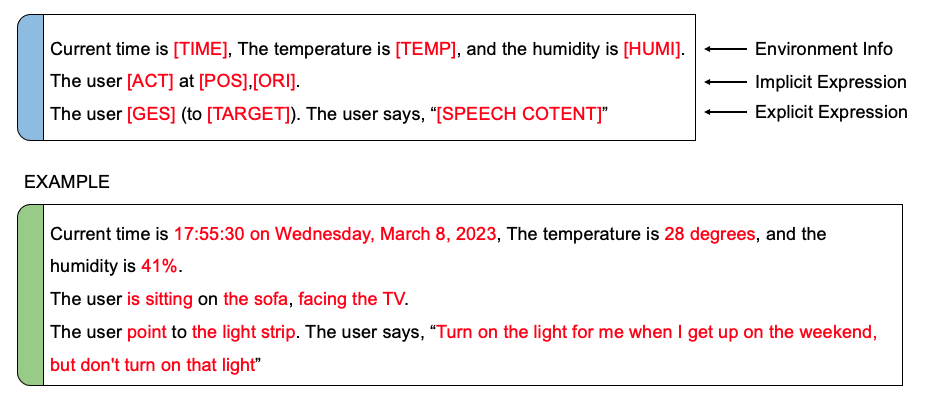}
\caption{Standardization of user expressions. The standardization contains the dynamic environment information and the description of the user behavior in natural language, which facilitates the inference of the subsequent model.}
\end{figure}
The environment provides a wealth of information, including device status and contextual factors such as weather and time. This information is crucial in the automation configuration process; the user provides much of it. We hope environmental information after standardization can supplement user input, enabling the system to reason more effectively about automation. For example, compared to "When the current weather occurs, turn on the light", users prefer to say, "When it is rainy, turn on the light". Therefore weather is not essential and is likely to be included in the user's expression. However, users always implicitly provide time information, such as creating a rule to "Play music at this time every day." Based on statistical analysis, we have identified time, temperature, and humidity as key factors to consider.

The user's expression involves a combination of implicit gestural activities and explicit speech and gestures. The rule-related information includes the user's current position, posture/activity, gesture, the target of the gesture, and speech content. We standardize these factors as "The user [posture/activity] on [position],[orientation], [gesture] (towards  [target]), and says [speech cotent]." We assign natural language semantics to the parameters to ensure the validity of inference. For instance, a normalized description would be "The user sits on the sofa, points towards ceiling light, and says, "Turn on this light when I sit here." We use target information such as sofa and light instead of specific coordinate parameters, which is more valuable for inference. This approach is beneficial for the system to reason about semantics. The existing sensors can be configured to convert location or target data to natural language.

The normalization process ensures that all the collected implicit parameters are accounted for, even in complex rules. This level of normalization is sufficient to support the inference of most complex rules and provide the necessary information for parameter inference.

\subsection{Reasoning logic of automation rules}

We use the inference ability of large language models to complete the reasoning of complex automation rules for users. To ensure the accuracy and standardization of the inference results, we carry out the design of the prompt to ensure that the user's information can be accurately captured and converted into the automation parameters and the user's logical expressions can be accurately converted into the triggers and actions in the automation rules.

\subsubsection{Standardization of automation operation}
Our study specifically focuses on the behavior of users editing automation. We provide the large language model with normalized user expression information to facilitate the reasoning of automation rules described in natural language. To handle the complexity of these rules, we have standardized the model's output. In the standardization of automation rules, we divide the information into four parts. 
\begin{figure}[htbp]
\centering
\includegraphics[width=12cm]{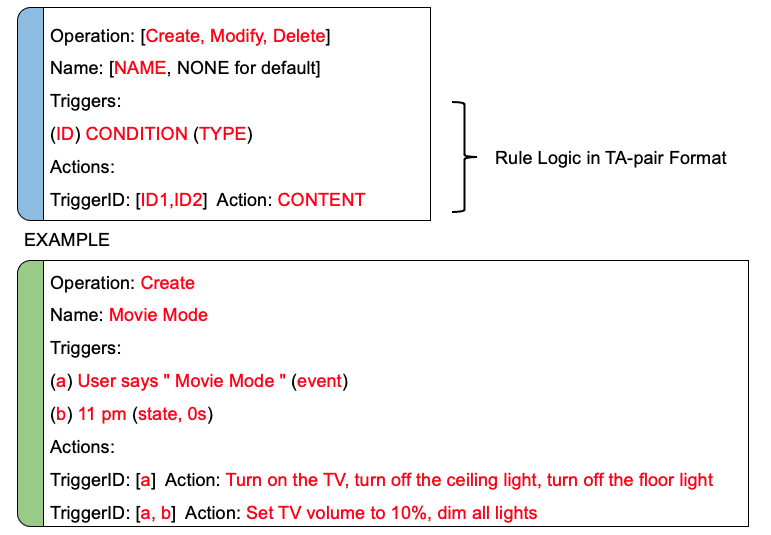}
\caption{Standardization of complex rules. In the rule logic relationship using TA-pair management, the trigger is responsible for recording all the relevant conditions. The actions are in the form of a group of records, each group recorded the id set of the trigger}
\end{figure}
The first part is the purpose of the user's operation, which can be creating, modifying, or deleting automation. Based on the user's behavior, the system will determine whether the user is creating new automation or modifying the logic of existing automation. Modifications include changing triggers, actions, or the relationship between them. 

The second part is the name of the automation rule. We have found that users often name their rules to facilitate their use and management. Rules with names generally have a default trigger, where the user directly says the rule's name to trigger the automation. 

The third part is the triggers of the automation rules. We have atomized the triggers to account for complex rules' temporal characteristics and branches. This means that triggers are a collection of atomized conditions. Each atomic condition is divided into an event (triggered when it occurs) or a state (triggered after a fixed time) according to its triggering mode. 

The fourth part is the actions. We designed the timing and branching for actions. For timing, we strictly define the order of execution, which must be arranged correctly. Suppose there is a delay in the execution process, such as turning on the air conditioner after five minutes. In that case, a timer will be inserted into the action sequence. We group actions and mark the trigger corresponding to the group to cope with branching situations. We have created a TA-pair concept, where each pair contains a trigger and action set. The action is set off when the trigger set conditions are satisfied.
We describe the output in the prompt, explain the rationale in the detail section, and finally reinforce it with several examples.

The standardized automation rules can cope with complex logical situations and allow users to understand the logic intuitively for their subsequent editing and utility.

\subsubsection{Prompt design for rule inference}
To improve the ability of the large language model to generate outputs that match the expectations, we focused on refining the design of the prompt. We divided the prompt into four segments: the standard form of automation, the details about automation generation, scenario information, and cases. Standardized user expressions will be provided to the model together with the prompts. The model will output the automation rule in the form we set.
\begin{figure}[htbp]
\centering
\includegraphics[width=14cm]{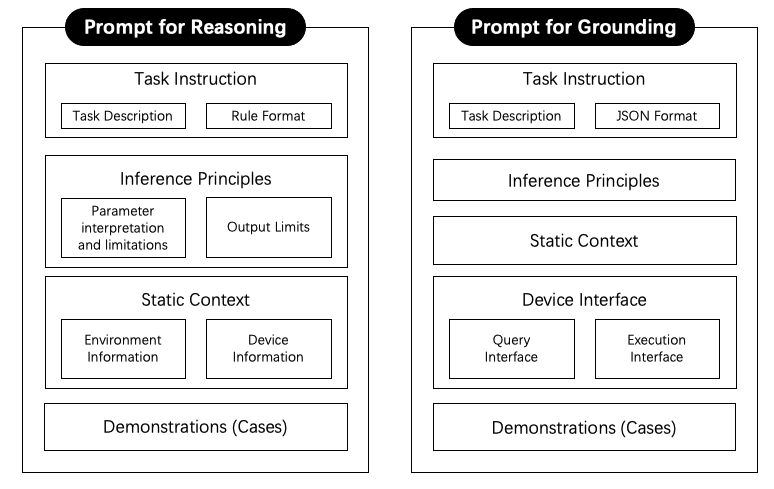}
\caption{Structure of prompt. The Reasoning and Grounding sections use a similar prompt design structure, starting with a formalization of both the output section and the inference constraints, giving the required static information in the middle section according to the requirements, and finally reinforcing LLM's understanding of the task and inference with examples.}
\end{figure}
The initial sections of the prompt serve to clarify and establish the model's output. We first outlined the target context and described the standardized output format. Additional information concerning the output content is provided, including details about the required components and parameter default values. We give some instructions to ensure that the output is limited to only the information relevant to the rule's operation to avoid extraneous or disruptive data.

We include scenario information about the device and environment in the prompt, including layout and device location details. This part of the environment information is also used for automated parametric reasoning. Since the layout and location typically remain constant regardless of user input, we have included them in the prompt to streamline the standardized user input.

At the end of the prompt, we provide ten illustrative examples. These examples encompass a range of operations, including creating, modifying, and deleting entities and rules with auxiliary parameters such as time, space, and user behavior. These examples include rules with default triggers for rule names, branching cases, and rules with a temporal order of triggers and actions. By including these examples, we aim to help the larger model understand the standardized format of input and output, as well as the logical relationships between them.

\subsection{Grounding automation rules}
Considering that the existing end-to-end reasoning capability of the LLM is limited and cannot well complete the one-step inference from the user's natural input to the available automation rules, we propose a split between reasoning and deployment capabilities. This method improves user controllability and interactivity. Users can adjust the inferred natural language rules to meet expectations. When problems arise, they can determine whether the cause is logic understanding or rule deployment based on the model's output. The input of the grounding part is the natural language rules generated by the previous model, and the output is the corresponding environment and device interface to help the system complete the deployment of automation rules.

\subsubsection{TA pair design for automation}
To ensure that complex automation rules are applicable in real-world scenarios, we utilized the powerful reasoning capabilities of a large language model. Our approach involved using the TA pair design to standardize automation rules, ensuring complex rules' validity. Each complex rule can be broken down into multiple TA pairs. Each TApair represents a simple automation rule consisting of triggers and actions. The actions are executed in order when all the triggers are met. This approach allows for retaining the completion semantics of complex rules without increasing management costs. It also aligns with user habits and makes it easier for users to understand and edit rules. The reasoning part handles the splitting of complex rules into TA pairs, while the grounding part converts natural language to an executable interface and selects the appropriate interface and parameters for the trigger and action.
\begin{figure}[htbp]
\centering
\includegraphics[width=12cm]{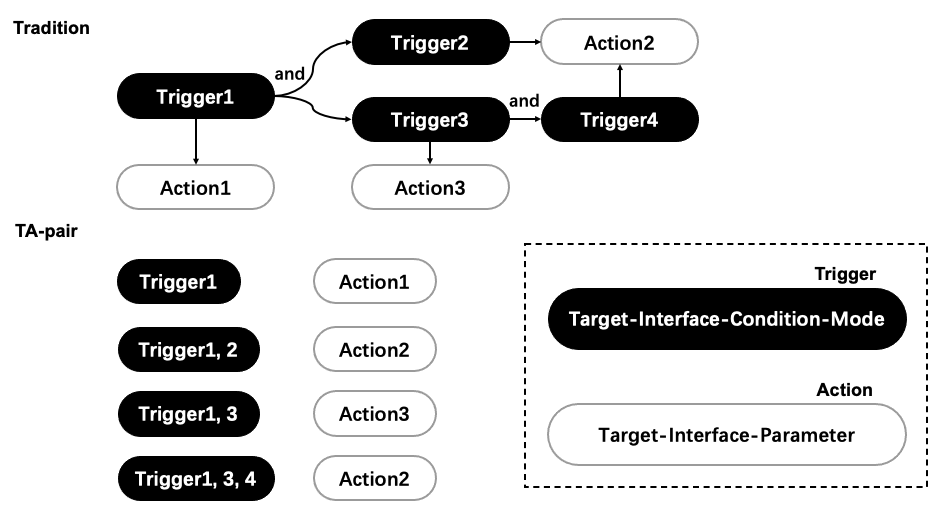}
\caption{TA-pair format Automation Rules. Compared with traditional branching rules, the TA-pair form of rules splits complex branches into intuitive, simple rules, which can facilitate users to understand and modify the logic of the rules and also provide convenience for the execution side of the system.}
\end{figure}
After analyzing the form of trigger and action, we have identified a tuple that can be used to describe them. For triggers, we have specified a quadruple consisting of "target-interface-condition-mode". This allows the system to determine whether a device or environmental information meets the conditions. We also labeled each trigger with a mode(event or state) and added a description of the delayed trigger time for the state case. For example, "when the TV is turned on" can be described as "TV-switch-on-event", while "after the user leaves the living room for 10 minutes" can be described as "ActivitySensor-isThereUserActivity-false-state(10mins)".

Actions only require the target and the instruction to execute the behavior. We have identified a ternary group with a "target-interface-parameter" to describe actions. The timing is solved by inserting a timer in the inference part. For example, "turn on the air conditioner for 10 minutes" becomes the action set "turn on the air conditioner, wait 10mins, turn off the air conditioner" after reasoning, and further grounding as "air conditioner-switch-on, timer-wait-10mins, air conditioner-switch-off." The Grounding part of the process converts the natural language rules into tuples and feeds them to the automation manager. 

\subsubsection{Prompt design for grounding}

The grounding part of our prompt design follows a similar structure to the inference part. We provide a normalized description of the output and impose certain constraints related to the output. Specifically, we propose using TApair and require the trigger and action to be in a tuple form to ensure consistency in the output. Additionally, we require the large language model to provide responses in JSON format for easy parsing and to avoid irrelevant content.

Layout information's role in grounding differs from that in reasoning, as it is used to determine the executability of a rule. The grounding process checks the feasibility of the specific target device and the rule, using the layout information to confirm the executability. In contrast, the inference part focuses on the logic of the general rule and may encounter issues with incomplete parameters or non-executable functions.

Interface information serves as a guide for grounding to generate instructions. We categorize device interfaces into queries and operations based on the correspondence between trigger and action. For each interface, we provide the corresponding target, parameters, and a natural language description of its function. In the case of query interfaces, we also provide the possible return values.

We conclude the prompt with an example of the conversion from natural language rules to JSON format rules to help the big language model understand the grounding process.

\subsection{Interaction with user}
The inference process relies on the user's natural expressions as input, and the system's reasoning and grounding generate automated rules as output. However, the rules may not be accurate due to limitations in both the model's reasoning ability and the user's expression. To improve the controllability of the process, we have incorporated system feedback and user interaction into the inference result of the model. This allows for refinement and adjustment of the generated rules, ensuring greater accuracy and reliability in the overall output.

\subsubsection{Optimized inference results}
In rule logic reasoning, we use the model to generate rules in natural language form. However, these rules may contain missing or incorrect parameters due to the potential ambiguity of user expressions and the limitations of the model's inference capabilities. To ensure user understanding, we present the natural language rules. Suppose the user finds that the inference result differs from their expectation. They can modify the rules using natural expressions or directly edit the natural language rules. This refinement allows for more accurate and practical rule-based reasoning.

\subsubsection{Help with rule deployment}
User uncertainty about device capabilities and names can lead to challenges in deploying rules. In some cases, rules may not be generated due to unsupported features or unclear parameters of the target device. 
To address this issue, we have implemented a feature that allows the model to output the reason for rule deployment failure. This provides the user with a specific reason why the rule is not executable and prompts them to clarify the parameters or modify the rule content. Users can modify the natural expression or edit the natural language rules to make them more precise and increase the chances of successful rule deployment.

\begin{figure}[htbp]
\centering
\includegraphics[width=14cm]{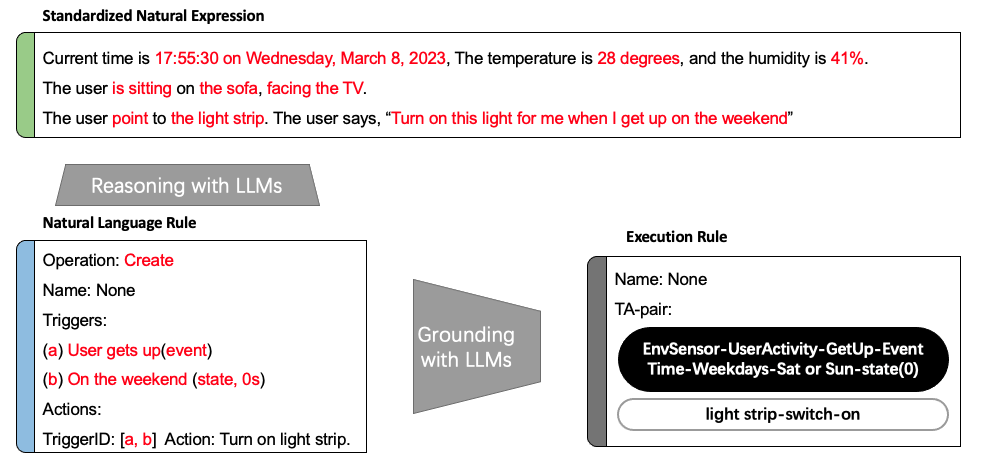}
\caption{Rule generation process. The process of one user expression from the formation of a normalized description to the final executable state is given in Fig. The model output of the two inference segments is shown in Fig.}
\end{figure}
The system achieves the process from natural user expression to complex automation reasoning by collaborating four components. Sensors in the environment first capture the user behavior and input, which is normalized to contain the necessary information for inference rules. The system sends the normalized user multimodal information to the reasoning part for the first inference, which uses the user input and the environment information to infer the rule logic. It processes the user's fuzzy expressions and redundant expressions to remove unnecessary information and complete the lost information. The first stage of reasoning obtains the automation rules of the natural language, which is fed back to the user, and this step does not block the subsequent reasoning process. The natural language rules continue to be sent to the grounding part for the second step of reasoning with deployable rules. The second stage of reasoning uses the interface information in the environment to determine the executability of the rules. It generates the files in JSON format to describe the interfaces and parameters associated with the rules and sends them to the automation management system. If there is an imperfect interface, missing functions, or uncertain parameters in the second reasoning stage, the problems will be described in the JSON. At the same time, the error causes will be recorded when the system parses the JSON file, and the problem in the second stage will also be fed back to the user. After getting feedback on the rule logic and deployment problems, users can finish the final automation generation by further interaction with the system or directly editing the rules.
The system achieves the process from natural user expression to complex automation reasoning by collaborating four components. Sensors in the environment first capture the user behavior and input, which is normalized to contain the necessary information for inference rules. The system sends the normalized user multimodal information to the reasoning part for the first inference, which uses the user input and the environment information to infer the rule logic. It processes the user's fuzzy expressions and redundant expressions to remove unnecessary information and complete the lost information. The first stage of reasoning obtains the automation rules of the natural language, which is fed back to the user, and this step does not block the subsequent reasoning process. The natural language rules continue to be sent to the grounding part for the second step of reasoning with deployable rules. The second stage of reasoning uses the interface information in the environment to determine the executability of the rules. It generates the files in JSON format to describe the interfaces and parameters associated with the rules and sends them to the automation management system. If there is an imperfect interface, missing functions, or uncertain parameters in the second reasoning stage, the problems will be described in the JSON. At the same time, the error causes will be recorded when the system parses the JSON file, and the problem in the second stage will also be fed back to the user. After getting feedback on the rule logic and deployment problems, users can finish the final automation generation by further interaction with the system or directly editing the rules.

\section{Performance Evaluation}
To verify the inference effect of the model, we first tested it on the existing dataset. Since the data in the dataset is offline, this part of the data is mainly used to test the inference capability of the system.

\subsection{Evaluation Indicators}
To assess the inference ability of the system at different stages, we designed three evaluation metrics:
\begin{itemize}
    \item The first is the \textbf{Inference Success Rate}. A successful rule is one that fully satisfies the user's intent at the time of configuration and works correctly in the environment. This requires that the rule completely and correctly parses the user's intent in both the reasoning and granting sections.
    \item The second is \textbf{Intent Consistency} in the reasoning section. We will illustrate intention consistency in terms of correctness and completeness. Correctness means that the inference result correctly completes the logical reasoning of the rule. Due to the ambiguity and diversity of natural language, the same user expression may be implemented using different rules. Our requirement for correctness exists mainly in the semantic layer, i.e., the rule's logic and intent must be consistent with the user expression. Completeness means that the inference rules can contain the necessary contents of the user's expression, such as understanding the parameters and completeness. Reasoning results that satisfy logical correctness may be incomplete, such as problems with the completeness of fuzzy or multimodal parameters. We consider a reasoning result to satisfy intent consistency when it satisfies both correctness and completeness.
    \item  The third is the \textbf{Feasibility} of the deployment part. Feasibility measures whether a rule can be correctly parsed into an executable automation rule. For a rule to be finally deployed, it must be executed correctly, and the objects in the rule must be correctly parsed into the environment's interfaces and invocation methods. We consider feasibility in terms of both executability and environmental conformance. There is a difference between feasibility and executability. Some rules here may have some errors in parsing but still generate executable rules. In addition, considering that the actual deployment of rules depends on the environment interface, if the environment interface does not satisfy the rule execution requirements, the executable rule itself cannot be generated. If the system gives the correct reason for non-executability, we still consider that it satisfies the environmental consistency. We consider a reasoning result satisfies feasibility when it satisfies both enforceability and environmental consistency.
\end{itemize}
These three evaluation metrics correspond to the different reasoning parts of the system and the reasoning ability of different complex rules. In addition, we will analyze the rules for inference errors, which can be caused by logical inference errors, parametric inference errors, omissions, or illusions.

\subsection{Complex task classification}
We discussed the sources of the complexity of the rule generation task in section\ref{challen}, so we classified the different data according to the complexity of the task as follows:
\begin{itemize}
    \item \textbf{Multi-parameter}: This rule type has many parameters and covers multiple trigger conditions or execution instructions. It is the basic type of complex rules.
    \item \textbf{Dynamic parameters}: There are dynamically variable parameters in the rules, which need to be judged according to the current scenario.
    \item \textbf{Multimodal parameters}: There is multimodal information fusion in the user expression, which needs to be combined with the context to complete the reasoning for the parameters.
    \item \textbf{Fuzzy expression}: There are fuzzy targets or parameters in the user's expression, which need to be combined with the context to complete the reasoning for the parameters
    \item \textbf{Redundant expressions}: The user has a lot of information in the expression that is irrelevant to the rule or needs to be ignored.
    \item \textbf{Complex branch}: There are branches between trigger and action in the execution logic of the rule, and the rule needs to execute different actions according to the trigger conditions
    \item \textbf{Time-related trigger}: There are time-related judgment requirements in the triggers of the rule, such as the delayed conditions.
    \item \textbf{Time-dependent action}: There are strict sequential or time-dependent execution conditions during the execution of the rule.
    \item \textbf{Combination}: The rules are a combination of the above-mentioned multiple complexity tasks
\end{itemize}

Different levels of complexity create challenges for different system capabilities, e.g., parameter-related rules have high requirements for the completeness of system reasoning while branching and timing-related rules rely on the correctness of system reasoning capabilities.

\subsection{Analysis of inference results}
We tested the inference of 205 data one by one and achieved a 91.7\% inference success rate from the results, among which 17 rules had problems in the generation process. For the two inference stages of the system, 96.59\% of intent consistency inference and 93.17\% of feasibility inference was achieved. In verifying the accuracy of grounding single-step inference, we manually corrected the natural language rules to ensure their intent consistency. Rule conflicts and interdependencies were not considered in the whole inference process. Overall, the system obtained good inference results. We will discuss the inference failure cases and the performance of different tasks in the following.

\subsubsection{Failure Cases Analysis}
We analyzed the rules of failure cases and wanted to discuss the limitations of the reasoning ability of this system. We attribute the failure cases to the following four categories according to the stages of reasoning failure and the causes of failure.

\begin{figure}[htbp]
\centering
\includegraphics[width=14cm]{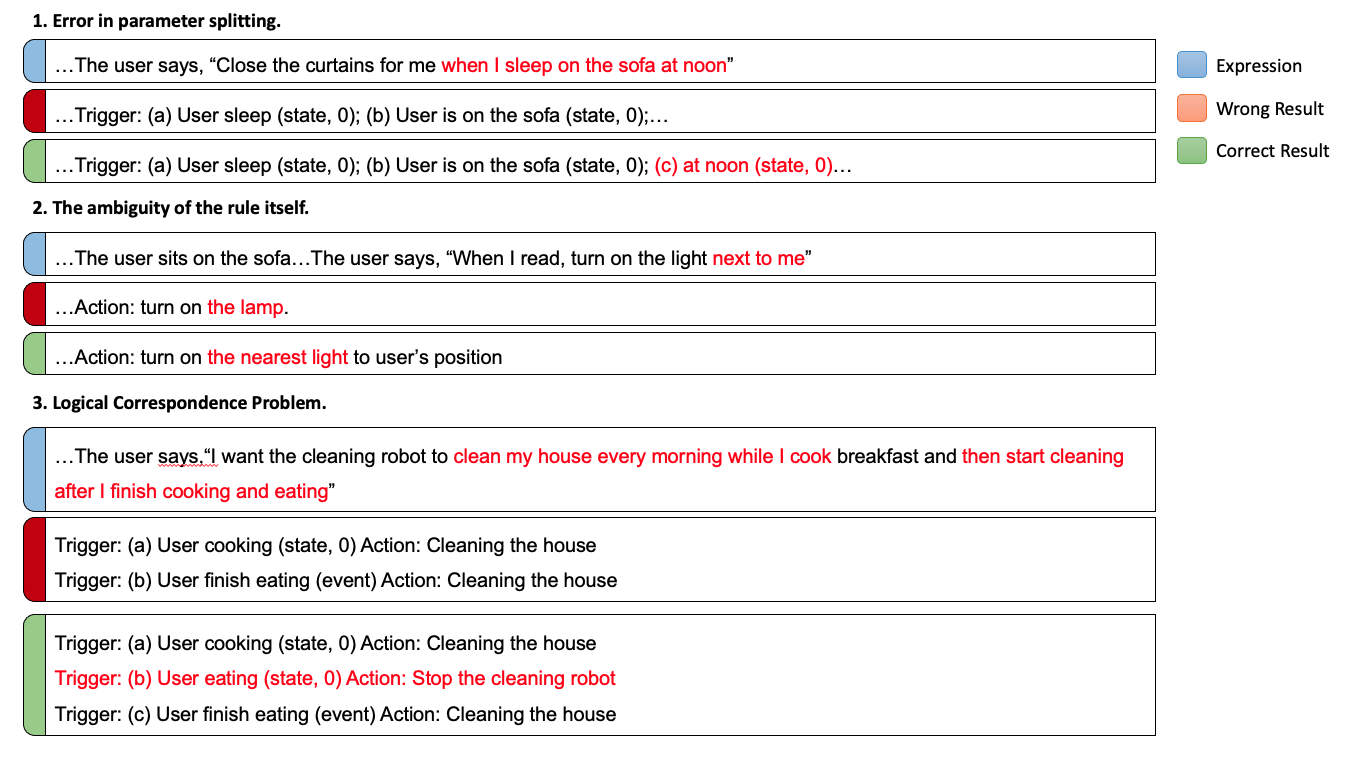}
\caption{Examples of three errors in the inference section. The figure shows the user's expression in blue, the result of incorrect reasoning in red, and the description of the correct rule in green.}
\end{figure}

\textbf{1. Error in parameter splitting.} Understanding and confirming the parameters is crucial for correctly inferring the automation rules. In the reasoning process, due to the limited understanding of the language model, some of the parameters are missing or misunderstood, mainly in the time-related parameters. For example, when the user expresses, "Close the curtains for me when I sleep on the sofa at noon", the system splits the trigger with only "the user sleep on the sofa", ignoring the time condition "at noon". Although this situation does not affect the execution of the rule in this example, the absence of the time condition will cause a possible rule mistrigger. The language model may also miss some of the information in the case of consecutive conditions, causing incompleteness of the rule. For example, the user's expression "Turn on movie mode when I enter this house at 5 pm on weekdays and weekends" covers many time periods. But the language model does not understand the parameters correctly when splitting the rule and identifies it as "5 pm on a weekday" and also "on the weekend", preventing the rule logic from being executed. The parameter errors in the inference are mainly in the case where the user's expression is too dense and the language model cannot accurately split it into the required number of parameters.

\textbf{2. The ambiguity of the rule itself.}
 The ambiguity and uncertainty in the description of the rules by the users make the rules potentially ambiguous. This ambiguity is mainly found in the reasoning of dynamic parameters. For example, if the user expresses, "When I read, turn on the light next to me", the "light next to me" refers to a lighting device that is dynamically related to the user's location. However, the system maps this parameter to a fixed device in the inference process, i.e., the light next to the user when he creates the rule. Another rule with similar logic, where the user's expression "Turn on the light over my head wherever I go." is correctly interpreted as a dynamic parameter. When the expression itself is ambiguous, the human understanding of the rule can be disambiguated based on the deep semantics of the rule or the relevant context. Understanding errors. In the above example of reading-light rules, most users would understand that turning on the lights is related to the act of reading, so the lights near the user should be turned on to help him/her read. But the large language model does not have this ability for the time being, which leads to the wrong understanding of expressions with ambiguity.

\textbf{3. Logical Correspondence Problem.}
In section\ref{challen}, we discuss the complex correspondence between trigger actions and the effect of temporal relations on rule complexity. The completing-ability of the large language model will be problematic when the user has default information in the expression of the temporal relations. The user expresses, "I want the cleaning robot to clean my house every morning while I cook breakfast and then start cleaning after I finish cooking and eating". In this rule, the user implicitly says that the robot should stop working during mealtime by using the keyword "then". The temporal constraint cannot be reasoned correctly. The missing actions and logic in the temporal sequence are difficult to be completed, which makes the correct automated reasoning challenging.

\textbf{4. The illusion of LLM.}
Since the large language model focuses on the complementary capabilities of expressions, some of the inference failures come from the illusions of the language model. This is mainly on the grounding stage. The language model may generate interfaces or return values based on natural language rules that do not exist or force execution to correspond to the wrong interface. For example, for the trigger "when a user enters the bedroom", the grounding part adds an [UserEnter] interface to the environment sensor during the reasoning process. For [user applaud], the model maps it to the user [snap finger]. This kind of forced complementation of the language model is the main reason for the failure of the grounding.

Considering the above four categories of failure, the reasoning ability of the large language model still has limitations. User expressions with more implicit information or relying much on situational understanding cannot be inferred correctly. However, due to the ambiguity of human expressions, it is difficult even for experimenters to understand all user behaviors correctly. Although we have achieved control over the language model by designing the prompt, there are still difficult problems in solving the illusion problem.
 
\subsubsection{Performance for different tasks}
We classified the rules according to their sources of complexity, where there are differences in the requirements of the system capabilities for different rules. Therefore we evaluated the system's performance for different complex rules by type. Among all complex rules, the multi-parameter complexity case is the most common (50/205) and the easiest to handle. These rules and the rules with fuzzy parameters (16/205) and redundant expressions (16/205) are reasoned successfully by the system. The user's logic is clearer in such rules, and the main difficulty lies in the reasoning of the parameters. The parameters of the fuzzy and redundant rules mainly come from understanding the user's words, so the success rate is higher.

\begin{table}[h]
  \centering
  \caption{Inference performance of different rules}
  \begin{tabular}{llllll}
    \toprule 
   {Rule Type}  & \multicolumn{2}{c}{Intent Consistency }&   \multicolumn{2}{c}{Feasibility}  &  {Success Rate} \\
      \cline{2-5} 
     & Correctness & Completeness&  Executability  & Env Conformance \\
    \midrule
  Multi-parameter(50) & 100 &100 &100  & 94.0 & 90.0\\
 Dynamic parameters(9) & 88.89 &88.89  &77.78 & 77.78  & 77.78\\
   Multimodal parameters(26) & 96.15 & 92.30 & 96.15 & 96.15 & 92.30 \\ 
Fuzzy expression(16) & 100 & 100 & 100  & 100  & 100  \\
Redundant expressions(16) & 100 & 100  &100 & 100  & 100\\
Complex branch(24) & 95.83 &95.83  &95.83 & 95.83  & 95.83\\
Time-related trigger(26) & 100 &96.15  &96.15 & 96.15  & 96.15\\
Time-dependent action(28) & 100 &96.42  &92.85 & 96.42  & 92.85\\
Combination(10) & 70.0 &70.0  &60.0 & 60.0  & 60.0\\

    \bottomrule
  \end{tabular}
   \caption*{The number in parentheses represents the number of corresponding rules}
  \label{tab:promptresult}
\end{table}

In contrast, the reasoning of four types of rules, branching rules (24/205), multimodal rules (26/205), Time-related trigger (26/205), and Time-related action (28/205), has some error cases. The reasoning for the completeness of multimodal rules is low. The user's parameter default in multimodal rules and the reliance on situational information are more likely to cause reasoning errors. The inference performance of dynamic parameter rules (9/205) and combination cases (10/205), which account for the smallest proportion, is the worst. The inference failure of dynamic rules mainly comes from the rules' ambiguity, while the inference results of combination cases are mainly limited by the complex rule logic and the diverse parameters involved in them. The inference of such rules needs to rely on a model with stronger inference capabilities.



\subsection{Interaction Evaluation}
In order to test the contribution of the interaction part to the inference results, we used Unity2018 software to build a smart living room environment in virtual reality and deployed 18 interactable devices. The size and location of all devices are consistent with the actual usage.
We used Microsoft's Azure and Oculus to capture the user's speech content, gestures, position, and orientation and converted them to the natural language description. These perception capabilities can be accomplished in real-world scenarios using state-of-art techniques.We build a Python server to communicate with ChatGPT and obtain the final inference results. 

We provide the user with an interactive feedback panel in virtual reality to display the natural language rules. The user can interact with the natural expressions or directly click on the panel to modify and submit rules. The final JSON results will be verified in the system to ensure the interfaces can be executed correctly.
\begin{figure}[htbp]
\centering
\includegraphics[width=14cm]{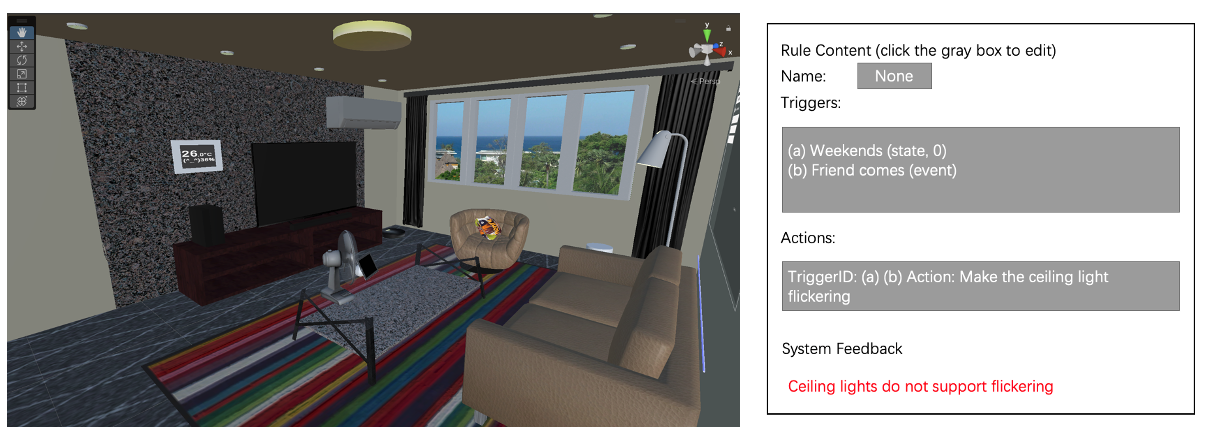}
\caption{Experimental environment in VR (Figure left) User feedback panel (Figure right). To ensure realism, the experimental environment specifies that the size and location of the equipment are the same as in reality. The user can edit the text in the rules between the experimental process to complete the modification of the results of the first inference stage.}
\end{figure}
We consider only the combination cases in the task because such rules depend more on the interaction. We will describe the rule's intent to the user and give the requirements of the expression and rule, and the user decides the specific content of the rule. Each user must complete a successful rule creation through interaction. The experimenter first introduces the user to the experiment content and system functionality and then guides the user through five rule-creation tasks.


We invited ten users (6 male, mean age 21.8,std=2.66) to participate in the experiment. Eight of these users did not have extensive experience using automated programming.
We obtained a total of 50 complex rule creation processes, of which 38 (76\%) were created without relying on subsequent interactions, 48 (94\%) were created within three rounds of input, and all interactions could be created within five rounds of interaction.
\begin{figure}[htbp]
\centering
\includegraphics[width=14cm]{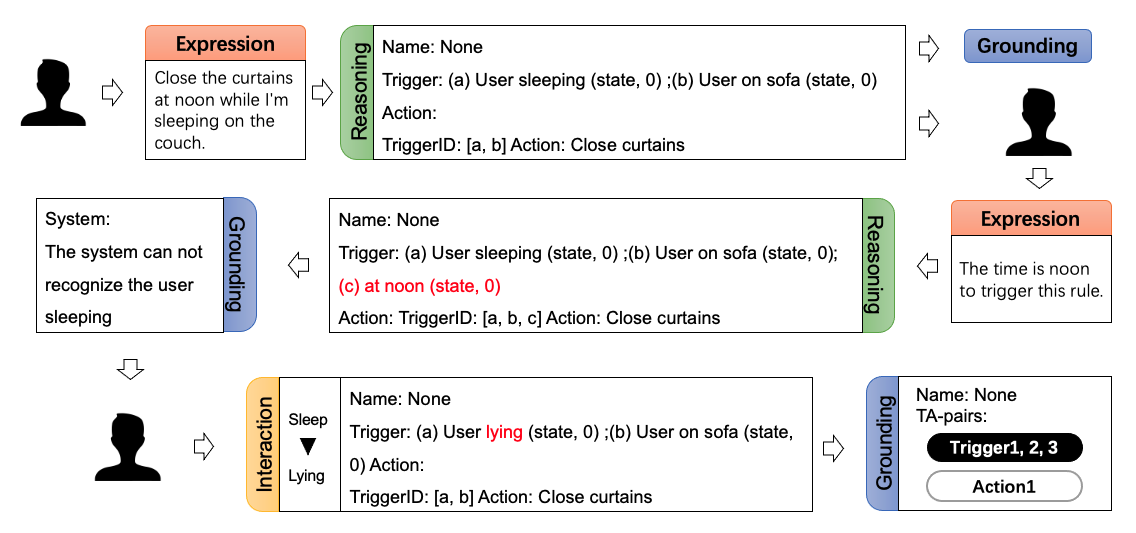}
\caption{Example of multiple rounds interaction to complete rule generation. The user provided information with natural expressions in the first round, and the reasoning result was returned to the user. The user found some parameters are missed and used natural expressions to give modification operations in the second round. In the second round, there are unattainable operations in the grounding part. The user directly modifies the natural language rules in the third round to re-grounding to obtain the final rules.}
\end{figure}
\textbf{User interaction preferences.} We observe that users choose different interaction schemes for modification when they find that the reasoning results are inconsistent with their expectations. When there are large deviations in the logic of the rule, users tend to re-express the rule using parameters and a clearer logic. When some of the parameters in the rule are misunderstood, the user will directly give the part of the rule or the parameter that needs to be modified. Only when the user cannot achieve correct creation through two rounds of interaction, he will use the panel to edit the rule for modification.

\textbf{User expression change.} An interesting phenomenon is that after a user's natural expression is not correctly recognized at a certain time, the ambiguity of all subsequent expressions decreases. After the user's pointing target was incorrectly identified, all subsequent expressions used language to describe the target device. Users constantly adjust their interaction with the system to trigger as few subsequent additional interactions as possible while ensuring the interaction is natural.

\section{Limitations and Discussion}





In this section, we discuss limitations, considerations, and potential future research directions related to our work.

\subsection{Dataset Size and Coverage}
In this work, we adopted a small-scale and single-scene dataset, with only 205 data samples of one smart home setting, for offline evaluation because to the best of our knowledge, the in-situ programming (ISP) dataset \cite{ISP} we used is the only existing dataset that contains complex configurations with multimodal input (voice, gesture, spatial context, etc.) collected from natural user interaction powered by an Wizard-of-Oz system. Although such a dataset contains high-quality in-situ programming data, the size is limited by the complexity of the configurations and the Wizard-of-Oz process. The limited size of dataset would lead to several drawbacks. For example, a turning-then-evaluation or cross-validation strategy for prompt turning and evaluation is not feasible due to insufficient samples. To overcome this problem, one strategy is to adopt data augmentation strategies such as paraphrasing and entity replacement \cite{feng2021survey} to extend the capability of dataset. The other strategy is to generate simulated data by various text generation techniques such as LLM \cite{gpt4}, template-based generation \cite{kale2020template}, and crowdsource human writing\cite{li2012crowdsourcing}. 

Also, we have noticed ISP dataset only contained data from one scene (the living room). Although we took the first step to validate the effectiveness of using large language to address complex automation problems, the evaluation is limited to the scene. Future work on investigating the performance of our algorithms in different scenes (e.g., kitchen, bedroom, office, etc.) is worthy of further research. Moreover, it is also worth investigating the scalability effect of our method brought by scene complexity (e.g., number of devices, different layouts, etc.).

\subsection{Interaction Modality Design}
AwareAuto mainly focused on addressing the problem of complex rule configuration in smart homes, which is the major part of in-situ programming. Currently in our implementation and evaluation study, users were instructed to perform immediate modification by either re-performing the configuration or manually modifying the rules by conversation or UI manipulation. However, a proper interaction design for retrieving, editing, and deleting certain rules is of equal importance to achieve a full-functional IoT agent for practical deployment of AwareAuto. Further research to incorporate additional prompt design and interaction design for retrieving, editing, and deleting rules is worthy of further investigation.

\subsection{Rule Management and Maintenance}
In our work, for the simplicity, all of the created rules were treated individually. We mainly focused on the semantics of individual rules themselves, while not imposing restriction on the relationship between different rules. For example, when the user says "Open the window when it is raining outside," and there is an existing rule indicating closing the window when it rains, the system should report a conflict on the newly created rule. To achieve this goal, it is essential to build a database or knowledge base for the management and maintenance of increasing rules. Such a rule management system should be capable for dealing with dependency, redundancy, conflict, and other inter-rule relationships, while giving the feedback to the frontend LLM agent. A possible solution of communication between the frontend LLM agent and the backend management system is through a tool-chain prompt design \cite{wu2022ai}.


\section{Conclusion}
This paper proposed \projectName{}, a LLM-based End-user programming (EUP) system for smart home automation. By combining LLMs inference, context-awareness, and real-time interaction, \projectName{} allows natural user expressions for complex automation tasks programming. We managed the challenges of reasoning from natural, flexible, ambiguous expressions, and grounding complex smart home contexts. We designed the prompts for reasoning and grounding where we defined and standardized essential time-spatial information for smart home automation reference. Through evaluation on the in-situ programming dataset, \projectName{} gains 91.7\% accuracy in matching user intentions and feasibility. \projectName{} filled the gap in EUP systems that balance natural user expression with complex, dynamic, and multi-modal automation rules. It presented the potential to incorporate LLMs into complex smart home tasks and improve naturalness and expressiveness for future end-user programming techniques.

\bibliographystyle{ACM-Reference-Format}
\bibliography{sample-base}


\begin{thebibliography}{79}


\ifx \showCODEN    \undefined \def \showCODEN     #1{\unskip}     \fi
\ifx \showDOI      \undefined \def \showDOI       #1{#1}\fi
\ifx \showISBNx    \undefined \def \showISBNx     #1{\unskip}     \fi
\ifx \showISBNxiii \undefined \def \showISBNxiii  #1{\unskip}     \fi
\ifx \showISSN     \undefined \def \showISSN      #1{\unskip}     \fi
\ifx \showLCCN     \undefined \def \showLCCN      #1{\unskip}     \fi
\ifx \shownote     \undefined \def \shownote      #1{#1}          \fi
\ifx \showarticletitle \undefined \def \showarticletitle #1{#1}   \fi
\ifx \showURL      \undefined \def \showURL       {\relax}        \fi
\providecommand\bibfield[2]{#2}
\providecommand\bibinfo[2]{#2}
\providecommand\natexlab[1]{#1}
\providecommand\showeprint[2][]{arXiv:#2}

\bibitem[ift(2022)]%
        {ifttt}
 \bibinfo{year}{2022}\natexlab{}.
\newblock \bibinfo{title}{IFTTT.}
\newblock
\newblock
\urldef\tempurl%
\url{https://ifttt.com}
\showURL{%
\tempurl}


\bibitem[gpt(2023)]%
        {gpt4}
 \bibinfo{year}{2023}\natexlab{}.
\newblock \bibinfo{title}{GPT4.}
\newblock
\newblock
\urldef\tempurl%
\url{https://openai.com/research/gpt-4}
\showURL{%
\tempurl}


\bibitem[nod(2023)]%
        {nodered}
 \bibinfo{year}{2023}\natexlab{}.
\newblock \bibinfo{title}{nodered.}
\newblock
\newblock
\urldef\tempurl%
\url{http://www.nodered.org}
\showURL{%
\tempurl}


\bibitem[Agadakos et~al\mbox{.}(2018)]%
        {agadakos2018butterfly}
\bibfield{author}{\bibinfo{person}{Ioannis Agadakos}, \bibinfo{person}{Gabriela
  Ciocarlie}, \bibinfo{person}{Bogdan Copos}, \bibinfo{person}{Tancrede
  Lepoint}, \bibinfo{person}{Ulf Lindqvist}, {and} \bibinfo{person}{Michael
  Locasto}.} \bibinfo{year}{2018}\natexlab{}.
\newblock \showarticletitle{Butterfly effect: causality from chaos in the IoT}.
  In \bibinfo{booktitle}{\emph{International Workshop on Security and Privacy
  for the Internet-of-Things}}. \bibinfo{pages}{26--30}.
\newblock


\bibitem[Ahn et~al\mbox{.}(2022)]%
        {ahn2022can}
\bibfield{author}{\bibinfo{person}{Michael Ahn}, \bibinfo{person}{Anthony
  Brohan}, \bibinfo{person}{Noah Brown}, \bibinfo{person}{Yevgen Chebotar},
  \bibinfo{person}{Omar Cortes}, \bibinfo{person}{Byron David},
  \bibinfo{person}{Chelsea Finn}, \bibinfo{person}{Keerthana Gopalakrishnan},
  \bibinfo{person}{Karol Hausman}, \bibinfo{person}{Alex Herzog},
  {et~al\mbox{.}}} \bibinfo{year}{2022}\natexlab{}.
\newblock \showarticletitle{Do as i can, not as i say: Grounding language in
  robotic affordances}.
\newblock \bibinfo{journal}{\emph{arXiv preprint arXiv:2204.01691}}
  (\bibinfo{year}{2022}).
\newblock


\bibitem[Ammari et~al\mbox{.}(2019)]%
        {ammari2019music}
\bibfield{author}{\bibinfo{person}{Tawfiq Ammari}, \bibinfo{person}{Jofish
  Kaye}, \bibinfo{person}{Janice~Y Tsai}, {and} \bibinfo{person}{Frank
  Bentley}.} \bibinfo{year}{2019}\natexlab{}.
\newblock \showarticletitle{Music, Search, and IoT: How People (Really) Use
  Voice Assistants.}
\newblock \bibinfo{journal}{\emph{ACM Trans. Comput. Hum. Interact.}}
  \bibinfo{volume}{26}, \bibinfo{number}{3} (\bibinfo{year}{2019}),
  \bibinfo{pages}{17--1}.
\newblock


\bibitem[Ardito et~al\mbox{.}(2020)]%
        {ardito2020user}
\bibfield{author}{\bibinfo{person}{Carmelo Ardito}, \bibinfo{person}{Giuseppe
  Desolda}, \bibinfo{person}{Rosa Lanzilotti}, \bibinfo{person}{Alessio
  Malizia}, \bibinfo{person}{Maristella Matera}, \bibinfo{person}{Paolo Buono},
  {and} \bibinfo{person}{Antonio Piccinno}.} \bibinfo{year}{2020}\natexlab{}.
\newblock \showarticletitle{User-defined semantics for the design of IoT
  systems enabling smart interactive experiences}.
\newblock \bibinfo{journal}{\emph{Personal and Ubiquitous Computing}}
  \bibinfo{volume}{24}, \bibinfo{number}{6} (\bibinfo{year}{2020}),
  \bibinfo{pages}{781--796}.
\newblock


\bibitem[Ariano et~al\mbox{.}(2022)]%
        {ariano2022smartphone}
\bibfield{author}{\bibinfo{person}{Raffaele Ariano}, \bibinfo{person}{Marco
  Manca}, \bibinfo{person}{Fabio Patern{\`o}}, {and} \bibinfo{person}{Carmen
  Santoro}.} \bibinfo{year}{2022}\natexlab{}.
\newblock \showarticletitle{Smartphone-based augmented reality for end-user
  creation of home automations}.
\newblock \bibinfo{journal}{\emph{Behaviour \& Information Technology}}
  (\bibinfo{year}{2022}), \bibinfo{pages}{1--17}.
\newblock


\bibitem[Barricelli et~al\mbox{.}(2019)]%
        {barricelli2019end}
\bibfield{author}{\bibinfo{person}{Barbara~Rita Barricelli},
  \bibinfo{person}{Fabio Cassano}, \bibinfo{person}{Daniela Fogli}, {and}
  \bibinfo{person}{Antonio Piccinno}.} \bibinfo{year}{2019}\natexlab{}.
\newblock \showarticletitle{End-user development, end-user programming and
  end-user software engineering: A systematic mapping study}.
\newblock \bibinfo{journal}{\emph{Journal of Systems and Software}}
  \bibinfo{volume}{149} (\bibinfo{year}{2019}), \bibinfo{pages}{101--137}.
\newblock


\bibitem[Barricelli and Fogli(2022)]%
        {barricelli2022exploring}
\bibfield{author}{\bibinfo{person}{Barbara~Rita Barricelli} {and}
  \bibinfo{person}{Daniela Fogli}.} \bibinfo{year}{2022}\natexlab{}.
\newblock \showarticletitle{Exploring the Reciprocal Influence of Artificial
  Intelligence and End-User Development}.
\newblock  (\bibinfo{year}{2022}).
\newblock


\bibitem[Barricelli et~al\mbox{.}(2022)]%
        {barricelli2022multi}
\bibfield{author}{\bibinfo{person}{Barbara~Rita Barricelli},
  \bibinfo{person}{Daniela Fogli}, \bibinfo{person}{Letizia Iemmolo}, {and}
  \bibinfo{person}{Angela Locoro}.} \bibinfo{year}{2022}\natexlab{}.
\newblock \showarticletitle{A multi-modal approach to creating routines for
  smart speakers}. In \bibinfo{booktitle}{\emph{Proceedings of the 2022
  International Conference on Advanced Visual Interfaces}}.
  \bibinfo{pages}{1--5}.
\newblock


\bibitem[Barricelli and Valtolina(2015)]%
        {barricelli2015designing}
\bibfield{author}{\bibinfo{person}{Barbara~Rita Barricelli} {and}
  \bibinfo{person}{Stefano Valtolina}.} \bibinfo{year}{2015}\natexlab{}.
\newblock \showarticletitle{Designing for end-user development in the internet
  of things}. In \bibinfo{booktitle}{\emph{International symposium on end user
  development}}. Springer, \bibinfo{pages}{9--24}.
\newblock


\bibitem[Betz et~al\mbox{.}(2021)]%
        {betz2021thinking}
\bibfield{author}{\bibinfo{person}{Gregor Betz}, \bibinfo{person}{Kyle
  Richardson}, {and} \bibinfo{person}{Christian Voigt}.}
  \bibinfo{year}{2021}\natexlab{}.
\newblock \showarticletitle{Thinking aloud: Dynamic context generation improves
  zero-shot reasoning performance of gpt-2}.
\newblock \bibinfo{journal}{\emph{arXiv preprint arXiv:2103.13033}}
  (\bibinfo{year}{2021}).
\newblock


\bibitem[Billard et~al\mbox{.}(2008)]%
        {billard2008robot}
\bibfield{author}{\bibinfo{person}{Aude Billard}, \bibinfo{person}{Sylvain
  Calinon}, \bibinfo{person}{Ruediger Dillmann}, {and} \bibinfo{person}{Stefan
  Schaal}.} \bibinfo{year}{2008}\natexlab{}.
\newblock \showarticletitle{Robot programming by demonstration}.
\newblock In \bibinfo{booktitle}{\emph{Springer handbook of robotics}}.
  \bibinfo{publisher}{Springer}, \bibinfo{pages}{1371--1394}.
\newblock


\bibitem[Bommasani et~al\mbox{.}(2021)]%
        {bommasani2021opportunities}
\bibfield{author}{\bibinfo{person}{Rishi Bommasani}, \bibinfo{person}{Drew~A
  Hudson}, \bibinfo{person}{Ehsan Adeli}, \bibinfo{person}{Russ Altman},
  \bibinfo{person}{Simran Arora}, \bibinfo{person}{Sydney von Arx},
  \bibinfo{person}{Michael~S Bernstein}, \bibinfo{person}{Jeannette Bohg},
  \bibinfo{person}{Antoine Bosselut}, \bibinfo{person}{Emma Brunskill},
  {et~al\mbox{.}}} \bibinfo{year}{2021}\natexlab{}.
\newblock \showarticletitle{On the opportunities and risks of foundation
  models}.
\newblock \bibinfo{journal}{\emph{arXiv preprint arXiv:2108.07258}}
  (\bibinfo{year}{2021}).
\newblock


\bibitem[Brown et~al\mbox{.}(2020)]%
        {brown2020language}
\bibfield{author}{\bibinfo{person}{Tom Brown}, \bibinfo{person}{Benjamin Mann},
  \bibinfo{person}{Nick Ryder}, \bibinfo{person}{Melanie Subbiah},
  \bibinfo{person}{Jared~D Kaplan}, \bibinfo{person}{Prafulla Dhariwal},
  \bibinfo{person}{Arvind Neelakantan}, \bibinfo{person}{Pranav Shyam},
  \bibinfo{person}{Girish Sastry}, \bibinfo{person}{Amanda Askell},
  {et~al\mbox{.}}} \bibinfo{year}{2020}\natexlab{}.
\newblock \showarticletitle{Language models are few-shot learners}.
\newblock \bibinfo{journal}{\emph{Advances in neural information processing
  systems}}  \bibinfo{volume}{33} (\bibinfo{year}{2020}),
  \bibinfo{pages}{1877--1901}.
\newblock


\bibitem[Caivano et~al\mbox{.}(2018)]%
        {caivano2018supporting}
\bibfield{author}{\bibinfo{person}{Danilo Caivano}, \bibinfo{person}{Daniela
  Fogli}, \bibinfo{person}{Rosa Lanzilotti}, \bibinfo{person}{Antonio
  Piccinno}, {and} \bibinfo{person}{Fabio Cassano}.}
  \bibinfo{year}{2018}\natexlab{}.
\newblock \showarticletitle{Supporting end users to control their smart home:
  design implications from a literature review and an empirical investigation}.
\newblock \bibinfo{journal}{\emph{Journal of Systems and Software}}
  \bibinfo{volume}{144} (\bibinfo{year}{2018}), \bibinfo{pages}{295--313}.
\newblock


\bibitem[Chen et~al\mbox{.}(2022)]%
        {chen2022open}
\bibfield{author}{\bibinfo{person}{Boyuan Chen}, \bibinfo{person}{Fei Xia},
  \bibinfo{person}{Brian Ichter}, \bibinfo{person}{Kanishka Rao},
  \bibinfo{person}{Keerthana Gopalakrishnan}, \bibinfo{person}{Michael~S Ryoo},
  \bibinfo{person}{Austin Stone}, {and} \bibinfo{person}{Daniel Kappler}.}
  \bibinfo{year}{2022}\natexlab{}.
\newblock \showarticletitle{Open-vocabulary queryable scene representations for
  real world planning}.
\newblock \bibinfo{journal}{\emph{arXiv preprint arXiv:2209.09874}}
  (\bibinfo{year}{2022}).
\newblock


\bibitem[Chin et~al\mbox{.}(2006)]%
        {chin2006end}
\bibfield{author}{\bibinfo{person}{Jeannette Shiaw-Yuan Chin},
  \bibinfo{person}{Victor Callaghan}, {and} \bibinfo{person}{Graham Clarke}.}
  \bibinfo{year}{2006}\natexlab{}.
\newblock \showarticletitle{An End-User Programming Paradigm for Pervasive
  Computing Applications.}. In \bibinfo{booktitle}{\emph{ICPS}},
  Vol.~\bibinfo{volume}{6}. \bibinfo{pages}{325--328}.
\newblock


\bibitem[Clark et~al\mbox{.}(2016)]%
        {clark2016towards}
\bibfield{author}{\bibinfo{person}{Meghan Clark}, \bibinfo{person}{Prabal
  Dutta}, {and} \bibinfo{person}{Mark~W Newman}.}
  \bibinfo{year}{2016}\natexlab{}.
\newblock \showarticletitle{Towards a natural language programming interface
  for smart homes}. In \bibinfo{booktitle}{\emph{Proceedings of the 2016 ACM
  International Joint Conference on Pervasive and Ubiquitous Computing:
  Adjunct}}. \bibinfo{pages}{49--52}.
\newblock


\bibitem[Corno et~al\mbox{.}(2020)]%
        {corno2020heytap}
\bibfield{author}{\bibinfo{person}{Fulvio Corno}, \bibinfo{person}{Luigi
  De~Russis}, {and} \bibinfo{person}{Alberto Monge~Roffarello}.}
  \bibinfo{year}{2020}\natexlab{}.
\newblock \showarticletitle{HeyTAP: Bridging the Gaps Between Users' Needs and
  Technology in IF-THEN Rules via Conversation}. In
  \bibinfo{booktitle}{\emph{Proceedings of the International Conference on
  Advanced Visual Interfaces}}. \bibinfo{pages}{1--9}.
\newblock


\bibitem[Corno et~al\mbox{.}(2021)]%
        {corno2021users}
\bibfield{author}{\bibinfo{person}{Fulvio Corno}, \bibinfo{person}{Luigi
  De~Russis}, {and} \bibinfo{person}{Alberto Monge~Roffarello}.}
  \bibinfo{year}{2021}\natexlab{}.
\newblock \showarticletitle{From users’ intentions to if-then rules in the
  internet of things}.
\newblock \bibinfo{journal}{\emph{ACM Transactions on Information Systems
  (TOIS)}} \bibinfo{volume}{39}, \bibinfo{number}{4} (\bibinfo{year}{2021}),
  \bibinfo{pages}{1--33}.
\newblock


\bibitem[Corno et~al\mbox{.}(2019)]%
        {highseman}
\bibfield{author}{\bibinfo{person}{Fulvio Corno}, \bibinfo{person}{Luigi
  De~Russis}, {and} \bibinfo{person}{Alberto~Monge Roffarello}.}
  \bibinfo{year}{2019}\natexlab{}.
\newblock \showarticletitle{A high-level semantic approach to end-user
  development in the Internet of Things}.
\newblock \bibinfo{journal}{\emph{International Journal of Human-Computer
  Studies}}  \bibinfo{volume}{125} (\bibinfo{year}{2019}),
  \bibinfo{pages}{41--54}.
\newblock


\bibitem[Desolda et~al\mbox{.}(2017)]%
        {desolda2017empowering}
\bibfield{author}{\bibinfo{person}{Giuseppe Desolda}, \bibinfo{person}{Carmelo
  Ardito}, {and} \bibinfo{person}{Maristella Matera}.}
  \bibinfo{year}{2017}\natexlab{}.
\newblock \showarticletitle{Empowering end users to customize their smart
  environments: model, composition paradigms, and domain-specific tools}.
\newblock \bibinfo{journal}{\emph{ACM Transactions on Computer-Human
  Interaction (TOCHI)}} \bibinfo{volume}{24}, \bibinfo{number}{2}
  (\bibinfo{year}{2017}), \bibinfo{pages}{1--52}.
\newblock


\bibitem[Devlin et~al\mbox{.}(2018)]%
        {devlin2018bert}
\bibfield{author}{\bibinfo{person}{Jacob Devlin}, \bibinfo{person}{Ming-Wei
  Chang}, \bibinfo{person}{Kenton Lee}, {and} \bibinfo{person}{Kristina
  Toutanova}.} \bibinfo{year}{2018}\natexlab{}.
\newblock \showarticletitle{Bert: Pre-training of deep bidirectional
  transformers for language understanding}.
\newblock \bibinfo{journal}{\emph{arXiv preprint arXiv:1810.04805}}
  (\bibinfo{year}{2018}).
\newblock


\bibitem[Dey et~al\mbox{.}(2004)]%
        {dey2004cappella}
\bibfield{author}{\bibinfo{person}{Anind~K Dey}, \bibinfo{person}{Raffay
  Hamid}, \bibinfo{person}{Chris Beckmann}, \bibinfo{person}{Ian Li}, {and}
  \bibinfo{person}{Daniel Hsu}.} \bibinfo{year}{2004}\natexlab{}.
\newblock \showarticletitle{a CAPpella: programming by demonstration of
  context-aware applications}. In \bibinfo{booktitle}{\emph{Proceedings of the
  SIGCHI conference on Human factors in computing systems}}.
  \bibinfo{pages}{33--40}.
\newblock


\bibitem[Feng et~al\mbox{.}(2021)]%
        {feng2021survey}
\bibfield{author}{\bibinfo{person}{Steven~Y Feng}, \bibinfo{person}{Varun
  Gangal}, \bibinfo{person}{Jason Wei}, \bibinfo{person}{Sarath Chandar},
  \bibinfo{person}{Soroush Vosoughi}, \bibinfo{person}{Teruko Mitamura}, {and}
  \bibinfo{person}{Eduard Hovy}.} \bibinfo{year}{2021}\natexlab{}.
\newblock \showarticletitle{A survey of data augmentation approaches for NLP}.
\newblock \bibinfo{journal}{\emph{arXiv preprint arXiv:2105.03075}}
  (\bibinfo{year}{2021}).
\newblock


\bibitem[Funk et~al\mbox{.}(2018)]%
        {funk2018addressing}
\bibfield{author}{\bibinfo{person}{Mathias Funk}, \bibinfo{person}{Lin-Lin
  Chen}, \bibinfo{person}{Shao-Wen Yang}, {and} \bibinfo{person}{Yen-Kuang
  Chen}.} \bibinfo{year}{2018}\natexlab{}.
\newblock \showarticletitle{Addressing the need to capture scenarios,
  intentions and preferences: Interactive intentional programming in the smart
  home}.
\newblock \bibinfo{journal}{\emph{International Journal of Design}}
  \bibinfo{volume}{12}, \bibinfo{number}{1} (\bibinfo{year}{2018}),
  \bibinfo{pages}{53--66}.
\newblock


\bibitem[Ghiani et~al\mbox{.}(2017)]%
        {ghiani2017personalization}
\bibfield{author}{\bibinfo{person}{Giuseppe Ghiani}, \bibinfo{person}{Marco
  Manca}, \bibinfo{person}{Fabio Patern{\`o}}, {and} \bibinfo{person}{Carmen
  Santoro}.} \bibinfo{year}{2017}\natexlab{}.
\newblock \bibinfo{journal}{\emph{ACM Transactions on Computer-Human
  Interaction (TOCHI)}} \bibinfo{volume}{24}, \bibinfo{number}{2}
  (\bibinfo{year}{2017}), \bibinfo{pages}{1--33}.
\newblock


\bibitem[Gordon and Harel(2014)]%
        {gordon2014steps}
\bibfield{author}{\bibinfo{person}{Michal Gordon} {and} \bibinfo{person}{David
  Harel}.} \bibinfo{year}{2014}\natexlab{}.
\newblock \showarticletitle{Steps towards scenario-based programming with a
  natural language interface}.
\newblock In \bibinfo{booktitle}{\emph{From Programs to Systems. The Systems
  perspective in Computing}}. \bibinfo{publisher}{Springer},
  \bibinfo{pages}{129--144}.
\newblock


\bibitem[Gu et~al\mbox{.}(2022)]%
        {gu2022don}
\bibfield{author}{\bibinfo{person}{Yu Gu}, \bibinfo{person}{Xiang Deng}, {and}
  \bibinfo{person}{Yu Su}.} \bibinfo{year}{2022}\natexlab{}.
\newblock \showarticletitle{Don't Generate, Discriminate: A Proposal for
  Grounding Language Models to Real-World Environments}.
\newblock \bibinfo{journal}{\emph{arXiv preprint arXiv:2212.09736}}
  (\bibinfo{year}{2022}).
\newblock


\bibitem[Huang and Cakmak(2015)]%
        {huang2015supporting}
\bibfield{author}{\bibinfo{person}{Justin Huang} {and} \bibinfo{person}{Maya
  Cakmak}.} \bibinfo{year}{2015}\natexlab{}.
\newblock \showarticletitle{Supporting mental model accuracy in trigger-action
  programming}. In \bibinfo{booktitle}{\emph{Proceedings of the 2015 ACM
  International Joint Conference on Pervasive and Ubiquitous Computing}}.
  \bibinfo{pages}{215--225}.
\newblock


\bibitem[Huang et~al\mbox{.}(2016)]%
        {huang2016instructablecrowd}
\bibfield{author}{\bibinfo{person}{Ting-Hao~Kenneth Huang},
  \bibinfo{person}{Amos Azaria}, {and} \bibinfo{person}{Jeffrey~P Bigham}.}
  \bibinfo{year}{2016}\natexlab{}.
\newblock \showarticletitle{Instructablecrowd: Creating if-then rules via
  conversations with the crowd}. In \bibinfo{booktitle}{\emph{Proceedings of
  the 2016 CHI Conference Extended Abstracts on Human Factors in Computing
  Systems}}. \bibinfo{pages}{1555--1562}.
\newblock


\bibitem[Huang et~al\mbox{.}(2022a)]%
        {huang2022language}
\bibfield{author}{\bibinfo{person}{Wenlong Huang}, \bibinfo{person}{Pieter
  Abbeel}, \bibinfo{person}{Deepak Pathak}, {and} \bibinfo{person}{Igor
  Mordatch}.} \bibinfo{year}{2022}\natexlab{a}.
\newblock \showarticletitle{Language models as zero-shot planners: Extracting
  actionable knowledge for embodied agents}. In
  \bibinfo{booktitle}{\emph{International Conference on Machine Learning}}.
  PMLR, \bibinfo{pages}{9118--9147}.
\newblock


\bibitem[Huang et~al\mbox{.}(2022b)]%
        {huang2022inner}
\bibfield{author}{\bibinfo{person}{Wenlong Huang}, \bibinfo{person}{Fei Xia},
  \bibinfo{person}{Ted Xiao}, \bibinfo{person}{Harris Chan},
  \bibinfo{person}{Jacky Liang}, \bibinfo{person}{Pete Florence},
  \bibinfo{person}{Andy Zeng}, \bibinfo{person}{Jonathan Tompson},
  \bibinfo{person}{Igor Mordatch}, \bibinfo{person}{Yevgen Chebotar},
  {et~al\mbox{.}}} \bibinfo{year}{2022}\natexlab{b}.
\newblock \showarticletitle{Inner monologue: Embodied reasoning through
  planning with language models}.
\newblock \bibinfo{journal}{\emph{arXiv preprint arXiv:2207.05608}}
  (\bibinfo{year}{2022}).
\newblock


\bibitem[Jakobi et~al\mbox{.}(2018)]%
        {jakobi2018evolving}
\bibfield{author}{\bibinfo{person}{Timo Jakobi}, \bibinfo{person}{Gunnar
  Stevens}, \bibinfo{person}{Nico Castelli}, \bibinfo{person}{Corinna
  Ogonowski}, \bibinfo{person}{Florian Schaub}, \bibinfo{person}{Nils Vindice},
  \bibinfo{person}{Dave Randall}, \bibinfo{person}{Peter Tolmie}, {and}
  \bibinfo{person}{Volker Wulf}.} \bibinfo{year}{2018}\natexlab{}.
\newblock \showarticletitle{Evolving needs in IoT control and accountability: A
  longitudinal study on smart home intelligibility}.
\newblock \bibinfo{journal}{\emph{Proceedings of the ACM on Interactive,
  Mobile, Wearable and Ubiquitous Technologies}} \bibinfo{volume}{2},
  \bibinfo{number}{4} (\bibinfo{year}{2018}), \bibinfo{pages}{1--28}.
\newblock


\bibitem[Janssen et~al\mbox{.}(2014)]%
        {Janssen2014VisualDM}
\bibfield{author}{\bibinfo{person}{Patrick Janssen}, \bibinfo{person}{Halil
  Erhan}, {and} \bibinfo{person}{Kian~Wee Chen}.}
  \bibinfo{year}{2014}\natexlab{}.
\newblock \showarticletitle{Visual Dataflow Modelling - Some thoughts on
  complexity}.
\newblock \bibinfo{journal}{\emph{Proceedings of the 32nd International
  Conference on Education and Research in Computer Aided Architectural Design
  in Europe (eCAADe) [Volume 2]}} (\bibinfo{year}{2014}).
\newblock


\bibitem[Kale and Rastogi(2020)]%
        {kale2020template}
\bibfield{author}{\bibinfo{person}{Mihir Kale} {and} \bibinfo{person}{Abhinav
  Rastogi}.} \bibinfo{year}{2020}\natexlab{}.
\newblock \showarticletitle{Template guided text generation for task-oriented
  dialogue}.
\newblock \bibinfo{journal}{\emph{arXiv preprint arXiv:2004.15006}}
  (\bibinfo{year}{2020}).
\newblock


\bibitem[Kang et~al\mbox{.}(2019)]%
        {kang2019minuet}
\bibfield{author}{\bibinfo{person}{Runchang Kang}, \bibinfo{person}{Anhong
  Guo}, \bibinfo{person}{Gierad Laput}, \bibinfo{person}{Yang Li}, {and}
  \bibinfo{person}{Xiang'Anthony' Chen}.} \bibinfo{year}{2019}\natexlab{}.
\newblock \showarticletitle{Minuet: Multimodal interaction with an internet of
  things}. In \bibinfo{booktitle}{\emph{Symposium on spatial user
  interaction}}. \bibinfo{pages}{1--10}.
\newblock


\bibitem[Kim and Ko(2022)]%
        {kim2022conversational}
\bibfield{author}{\bibinfo{person}{Sanghoon Kim} {and}
  \bibinfo{person}{In-Young Ko}.} \bibinfo{year}{2022}\natexlab{}.
\newblock \showarticletitle{A Conversational Approach for Modifying Service
  Mashups in IoT Environments}. In \bibinfo{booktitle}{\emph{CHI Conference on
  Human Factors in Computing Systems}}. \bibinfo{pages}{1--16}.
\newblock


\bibitem[Krishna et~al\mbox{.}(2022)]%
        {krishna2022socially}
\bibfield{author}{\bibinfo{person}{Ranjay Krishna}, \bibinfo{person}{Donsuk
  Lee}, \bibinfo{person}{Li Fei-Fei}, {and} \bibinfo{person}{Michael~S
  Bernstein}.} \bibinfo{year}{2022}\natexlab{}.
\newblock \showarticletitle{Socially situated artificial intelligence enables
  learning from human interaction}.
\newblock \bibinfo{journal}{\emph{Proceedings of the National Academy of
  Sciences}} \bibinfo{volume}{119}, \bibinfo{number}{39}
  (\bibinfo{year}{2022}), \bibinfo{pages}{e2115730119}.
\newblock


\bibitem[Lago et~al\mbox{.}(2021)]%
        {lago2021managing}
\bibfield{author}{\bibinfo{person}{Andr{\'e}~Sousa Lago},
  \bibinfo{person}{Jo{\~a}o~Pedro Dias}, {and} \bibinfo{person}{Hugo~Sereno
  Ferreira}.} \bibinfo{year}{2021}\natexlab{}.
\newblock \showarticletitle{Managing non-trivial internet-of-things systems
  with conversational assistants: A prototype and a feasibility experiment}.
\newblock \bibinfo{journal}{\emph{Journal of Computational Science}}
  \bibinfo{volume}{51} (\bibinfo{year}{2021}), \bibinfo{pages}{101324}.
\newblock


\bibitem[Lee et~al\mbox{.}(2013)]%
        {lee2013tangible}
\bibfield{author}{\bibinfo{person}{Jisoo Lee}, \bibinfo{person}{Luis
  Gardu{\~n}o}, \bibinfo{person}{Erin Walker}, {and} \bibinfo{person}{Winslow
  Burleson}.} \bibinfo{year}{2013}\natexlab{}.
\newblock \showarticletitle{A tangible programming tool for creation of
  context-aware applications}. In \bibinfo{booktitle}{\emph{Proceedings of the
  2013 ACM international joint conference on Pervasive and ubiquitous
  computing}}. \bibinfo{pages}{391--400}.
\newblock


\bibitem[Li et~al\mbox{.}(2012)]%
        {li2012crowdsourcing}
\bibfield{author}{\bibinfo{person}{Boyang Li}, \bibinfo{person}{Stephen
  Lee-Urban}, \bibinfo{person}{Darren~Scott Appling}, {and}
  \bibinfo{person}{Mark~O Riedl}.} \bibinfo{year}{2012}\natexlab{}.
\newblock \showarticletitle{Crowdsourcing narrative intelligence}.
\newblock \bibinfo{journal}{\emph{Advances in Cognitive systems}}
  \bibinfo{volume}{2}, \bibinfo{number}{1} (\bibinfo{year}{2012}).
\newblock


\bibitem[Li et~al\mbox{.}(2022)]%
        {li2022pre}
\bibfield{author}{\bibinfo{person}{Shuang Li}, \bibinfo{person}{Xavier Puig},
  \bibinfo{person}{Yilun Du}, \bibinfo{person}{Clinton Wang},
  \bibinfo{person}{Ekin Akyurek}, \bibinfo{person}{Antonio Torralba},
  \bibinfo{person}{Jacob Andreas}, {and} \bibinfo{person}{Igor Mordatch}.}
  \bibinfo{year}{2022}\natexlab{}.
\newblock \showarticletitle{Pre-trained language models for interactive
  decision-making}.
\newblock \bibinfo{journal}{\emph{arXiv preprint arXiv:2202.01771}}
  (\bibinfo{year}{2022}).
\newblock


\bibitem[Li et~al\mbox{.}(2018)]%
        {li2018appinite}
\bibfield{author}{\bibinfo{person}{Toby Jia-Jun Li}, \bibinfo{person}{Igor
  Labutov}, \bibinfo{person}{Xiaohan~Nancy Li}, \bibinfo{person}{Xiaoyi Zhang},
  \bibinfo{person}{Wenze Shi}, \bibinfo{person}{Wanling Ding},
  \bibinfo{person}{Tom~M Mitchell}, {and} \bibinfo{person}{Brad~A Myers}.}
  \bibinfo{year}{2018}\natexlab{}.
\newblock \showarticletitle{Appinite: A multi-modal interface for specifying
  data descriptions in programming by demonstration using natural language
  instructions}. In \bibinfo{booktitle}{\emph{2018 IEEE Symposium on Visual
  Languages and Human-Centric Computing (VL/HCC)}}. IEEE,
  \bibinfo{pages}{105--114}.
\newblock


\bibitem[Li et~al\mbox{.}(2017)]%
        {li2017programming}
\bibfield{author}{\bibinfo{person}{Toby Jia-Jun Li}, \bibinfo{person}{Yuanchun
  Li}, \bibinfo{person}{Fanglin Chen}, {and} \bibinfo{person}{Brad~A Myers}.}
  \bibinfo{year}{2017}\natexlab{}.
\newblock \showarticletitle{Programming IoT devices by demonstration using
  mobile apps}. In \bibinfo{booktitle}{\emph{International Symposium on End
  User Development}}. Springer, \bibinfo{pages}{3--17}.
\newblock


\bibitem[Liang et~al\mbox{.}(2022)]%
        {liang2022code}
\bibfield{author}{\bibinfo{person}{Jacky Liang}, \bibinfo{person}{Wenlong
  Huang}, \bibinfo{person}{Fei Xia}, \bibinfo{person}{Peng Xu},
  \bibinfo{person}{Karol Hausman}, \bibinfo{person}{Brian Ichter},
  \bibinfo{person}{Pete Florence}, {and} \bibinfo{person}{Andy Zeng}.}
  \bibinfo{year}{2022}\natexlab{}.
\newblock \showarticletitle{Code as policies: Language model programs for
  embodied control}.
\newblock \bibinfo{journal}{\emph{arXiv preprint arXiv:2209.07753}}
  (\bibinfo{year}{2022}).
\newblock


\bibitem[Lieberman et~al\mbox{.}(2006)]%
        {lieberman2006end}
\bibfield{author}{\bibinfo{person}{Henry Lieberman}, \bibinfo{person}{Fabio
  Patern{\`o}}, \bibinfo{person}{Markus Klann}, {and} \bibinfo{person}{Volker
  Wulf}.} \bibinfo{year}{2006}\natexlab{}.
\newblock \showarticletitle{End-user development: An emerging paradigm}.
\newblock In \bibinfo{booktitle}{\emph{End user development}}.
  \bibinfo{publisher}{Springer}, \bibinfo{pages}{1--8}.
\newblock


\bibitem[Liu and Shi(2023)]%
        {ISP}
\bibfield{author}{\bibinfo{person}{Liu} {and} \bibinfo{person}{Gao Yang
  Liang~Shi Shi, Yu}.} \bibinfo{year}{2023}\natexlab{}.
\newblock \showarticletitle{Understanding In-Situ Programming for Smart Home
  Automation}.
\newblock \bibinfo{journal}{\emph{Proceedings of the ACM on Interactive,
  Mobile, Wearable and Ubiquitous Technologies, Vol. 7, No. 2, Article 66.}}
  (\bibinfo{year}{2023}).
\newblock


\bibitem[Liu et~al\mbox{.}(2016)]%
        {liu2016latent}
\bibfield{author}{\bibinfo{person}{Chang Liu}, \bibinfo{person}{Xinyun Chen},
  \bibinfo{person}{Eui~Chul Shin}, \bibinfo{person}{Mingcheng Chen}, {and}
  \bibinfo{person}{Dawn Song}.} \bibinfo{year}{2016}\natexlab{}.
\newblock \showarticletitle{Latent attention for if-then program synthesis}.
\newblock \bibinfo{journal}{\emph{Advances in Neural Information Processing
  Systems}}  \bibinfo{volume}{29} (\bibinfo{year}{2016}).
\newblock


\bibitem[Liu et~al\mbox{.}(2021)]%
        {liu2021makes}
\bibfield{author}{\bibinfo{person}{Jiachang Liu}, \bibinfo{person}{Dinghan
  Shen}, \bibinfo{person}{Yizhe Zhang}, \bibinfo{person}{Bill Dolan},
  \bibinfo{person}{Lawrence Carin}, {and} \bibinfo{person}{Weizhu Chen}.}
  \bibinfo{year}{2021}\natexlab{}.
\newblock \showarticletitle{What Makes Good In-Context Examples for GPT-$3 $?}
\newblock \bibinfo{journal}{\emph{arXiv preprint arXiv:2101.06804}}
  (\bibinfo{year}{2021}).
\newblock


\bibitem[Lu et~al\mbox{.}(2021)]%
        {lu2021fantastically}
\bibfield{author}{\bibinfo{person}{Yao Lu}, \bibinfo{person}{Max Bartolo},
  \bibinfo{person}{Alastair Moore}, \bibinfo{person}{Sebastian Riedel}, {and}
  \bibinfo{person}{Pontus Stenetorp}.} \bibinfo{year}{2021}\natexlab{}.
\newblock \showarticletitle{Fantastically ordered prompts and where to find
  them: Overcoming few-shot prompt order sensitivity}.
\newblock \bibinfo{journal}{\emph{arXiv preprint arXiv:2104.08786}}
  (\bibinfo{year}{2021}).
\newblock


\bibitem[Mialon et~al\mbox{.}(2023)]%
        {mialon2023augmented}
\bibfield{author}{\bibinfo{person}{Gr{\'e}goire Mialon},
  \bibinfo{person}{Roberto Dess{\`\i}}, \bibinfo{person}{Maria Lomeli},
  \bibinfo{person}{Christoforos Nalmpantis}, \bibinfo{person}{Ram Pasunuru},
  \bibinfo{person}{Roberta Raileanu}, \bibinfo{person}{Baptiste Rozi{\`e}re},
  \bibinfo{person}{Timo Schick}, \bibinfo{person}{Jane Dwivedi-Yu},
  \bibinfo{person}{Asli Celikyilmaz}, {et~al\mbox{.}}}
  \bibinfo{year}{2023}\natexlab{}.
\newblock \showarticletitle{Augmented Language Models: a Survey}.
\newblock \bibinfo{journal}{\emph{arXiv preprint arXiv:2302.07842}}
  (\bibinfo{year}{2023}).
\newblock


\bibitem[Pane et~al\mbox{.}(2001)]%
        {pane2001studying}
\bibfield{author}{\bibinfo{person}{John~F Pane}, \bibinfo{person}{Brad~A
  Myers}, {et~al\mbox{.}}} \bibinfo{year}{2001}\natexlab{}.
\newblock \showarticletitle{Studying the language and structure in
  non-programmers' solutions to programming problems}.
\newblock \bibinfo{journal}{\emph{International Journal of Human-Computer
  Studies}} \bibinfo{volume}{54}, \bibinfo{number}{2} (\bibinfo{year}{2001}),
  \bibinfo{pages}{237--264}.
\newblock


\bibitem[Paschke(2006)]%
        {paschke2006eca}
\bibfield{author}{\bibinfo{person}{Adrian Paschke}.}
  \bibinfo{year}{2006}\natexlab{}.
\newblock \showarticletitle{ECA-RuleML: An approach combining ECA rules with
  temporal interval-based KR event/action logics and transactional update
  logics}.
\newblock \bibinfo{journal}{\emph{arXiv preprint cs/0610167}}
  (\bibinfo{year}{2006}).
\newblock


\bibitem[Qiao et~al\mbox{.}(2022)]%
        {qiao2022reasoning}
\bibfield{author}{\bibinfo{person}{Shuofei Qiao}, \bibinfo{person}{Yixin Ou},
  \bibinfo{person}{Ningyu Zhang}, \bibinfo{person}{Xiang Chen},
  \bibinfo{person}{Yunzhi Yao}, \bibinfo{person}{Shumin Deng},
  \bibinfo{person}{Chuanqi Tan}, \bibinfo{person}{Fei Huang}, {and}
  \bibinfo{person}{Huajun Chen}.} \bibinfo{year}{2022}\natexlab{}.
\newblock \showarticletitle{Reasoning with Language Model Prompting: A Survey}.
\newblock \bibinfo{journal}{\emph{arXiv preprint arXiv:2212.09597}}
  (\bibinfo{year}{2022}).
\newblock


\bibitem[Quirk et~al\mbox{.}(2015)]%
        {quirk2015language}
\bibfield{author}{\bibinfo{person}{Chris Quirk}, \bibinfo{person}{Raymond
  Mooney}, {and} \bibinfo{person}{Michel Galley}.}
  \bibinfo{year}{2015}\natexlab{}.
\newblock \showarticletitle{Language to code: Learning semantic parsers for
  if-this-then-that recipes}. In \bibinfo{booktitle}{\emph{Proceedings of the
  53rd Annual Meeting of the Association for Computational Linguistics and the
  7th International Joint Conference on Natural Language Processing (Volume 1:
  Long Papers)}}. \bibinfo{pages}{878--888}.
\newblock


\bibitem[Ray(2017)]%
        {ray2017survey}
\bibfield{author}{\bibinfo{person}{Partha~Pratim Ray}.}
  \bibinfo{year}{2017}\natexlab{}.
\newblock \showarticletitle{A survey on visual programming languages in
  internet of things}.
\newblock \bibinfo{journal}{\emph{Scientific Programming}}
  \bibinfo{volume}{2017} (\bibinfo{year}{2017}).
\newblock


\bibitem[Reynolds and McDonell(2021)]%
        {reynolds2021prompt}
\bibfield{author}{\bibinfo{person}{Laria Reynolds} {and} \bibinfo{person}{Kyle
  McDonell}.} \bibinfo{year}{2021}\natexlab{}.
\newblock \showarticletitle{Prompt programming for large language models:
  Beyond the few-shot paradigm}. In \bibinfo{booktitle}{\emph{Extended
  Abstracts of the 2021 CHI Conference on Human Factors in Computing Systems}}.
  \bibinfo{pages}{1--7}.
\newblock


\bibitem[Salovaara et~al\mbox{.}(2021)]%
        {salovaara2021programmable}
\bibfield{author}{\bibinfo{person}{Antti Salovaara}, \bibinfo{person}{Andrea
  Bellucci}, \bibinfo{person}{Andrea Vianello}, {and} \bibinfo{person}{Giulio
  Jacucci}.} \bibinfo{year}{2021}\natexlab{}.
\newblock \showarticletitle{Programmable Smart Home Toolkits Should Better
  Address Households’ Social Needs}. In \bibinfo{booktitle}{\emph{Proceedings
  of the 2021 CHI Conference on Human Factors in Computing Systems}}.
  \bibinfo{pages}{1--14}.
\newblock


\bibitem[Schick et~al\mbox{.}(2023)]%
        {schick2023toolformer}
\bibfield{author}{\bibinfo{person}{Timo Schick}, \bibinfo{person}{Jane
  Dwivedi-Yu}, \bibinfo{person}{Roberto Dess{\`\i}}, \bibinfo{person}{Roberta
  Raileanu}, \bibinfo{person}{Maria Lomeli}, \bibinfo{person}{Luke
  Zettlemoyer}, \bibinfo{person}{Nicola Cancedda}, {and}
  \bibinfo{person}{Thomas Scialom}.} \bibinfo{year}{2023}\natexlab{}.
\newblock \showarticletitle{Toolformer: Language models can teach themselves to
  use tools}.
\newblock \bibinfo{journal}{\emph{arXiv preprint arXiv:2302.04761}}
  (\bibinfo{year}{2023}).
\newblock


\bibitem[Seiger et~al\mbox{.}(2015)]%
        {seiger2015modelling}
\bibfield{author}{\bibinfo{person}{Ronny Seiger}, \bibinfo{person}{Christine
  Keller}, \bibinfo{person}{Florian Niebling}, {and} \bibinfo{person}{Thomas
  Schlegel}.} \bibinfo{year}{2015}\natexlab{}.
\newblock \showarticletitle{Modelling complex and flexible processes for smart
  cyber-physical environments}.
\newblock \bibinfo{journal}{\emph{Journal of Computational Science}}
  \bibinfo{volume}{10} (\bibinfo{year}{2015}), \bibinfo{pages}{137--148}.
\newblock


\bibitem[Shuster et~al\mbox{.}(2022)]%
        {shuster2022blenderbot}
\bibfield{author}{\bibinfo{person}{Kurt Shuster}, \bibinfo{person}{Jing Xu},
  \bibinfo{person}{Mojtaba Komeili}, \bibinfo{person}{Da Ju},
  \bibinfo{person}{Eric~Michael Smith}, \bibinfo{person}{Stephen Roller},
  \bibinfo{person}{Megan Ung}, \bibinfo{person}{Moya Chen},
  \bibinfo{person}{Kushal Arora}, \bibinfo{person}{Joshua Lane},
  {et~al\mbox{.}}} \bibinfo{year}{2022}\natexlab{}.
\newblock \showarticletitle{Blenderbot 3: a deployed conversational agent that
  continually learns to responsibly engage}.
\newblock \bibinfo{journal}{\emph{arXiv preprint arXiv:2208.03188}}
  (\bibinfo{year}{2022}).
\newblock


\bibitem[Singh et~al\mbox{.}(2022)]%
        {singh2022progprompt}
\bibfield{author}{\bibinfo{person}{Ishika Singh}, \bibinfo{person}{Valts
  Blukis}, \bibinfo{person}{Arsalan Mousavian}, \bibinfo{person}{Ankit Goyal},
  \bibinfo{person}{Danfei Xu}, \bibinfo{person}{Jonathan Tremblay},
  \bibinfo{person}{Dieter Fox}, \bibinfo{person}{Jesse Thomason}, {and}
  \bibinfo{person}{Animesh Garg}.} \bibinfo{year}{2022}\natexlab{}.
\newblock \showarticletitle{Progprompt: Generating situated robot task plans
  using large language models}.
\newblock \bibinfo{journal}{\emph{arXiv preprint arXiv:2209.11302}}
  (\bibinfo{year}{2022}).
\newblock


\bibitem[Song et~al\mbox{.}(2022)]%
        {song2022llm}
\bibfield{author}{\bibinfo{person}{Chan~Hee Song}, \bibinfo{person}{Jiaman Wu},
  \bibinfo{person}{Clayton Washington}, \bibinfo{person}{Brian~M Sadler},
  \bibinfo{person}{Wei-Lun Chao}, {and} \bibinfo{person}{Yu Su}.}
  \bibinfo{year}{2022}\natexlab{}.
\newblock \showarticletitle{Llm-planner: Few-shot grounded planning for
  embodied agents with large language models}.
\newblock \bibinfo{journal}{\emph{arXiv preprint arXiv:2212.04088}}
  (\bibinfo{year}{2022}).
\newblock


\bibitem[Stein et~al\mbox{.}(2016)]%
        {stein2016third}
\bibfield{author}{\bibinfo{person}{Martin Stein}, \bibinfo{person}{Alexander
  Boden}, \bibinfo{person}{Dominik Hornung}, \bibinfo{person}{Volker Wulf},
  \bibinfo{person}{In~Markus Garschall}, \bibinfo{person}{Theo Hamm},
  \bibinfo{person}{Claudia M{\"u}ller}, \bibinfo{person}{Katja Neureiter},
  \bibinfo{person}{Mar{\'e}n Schorch}, {and} \bibinfo{person}{Lex van Velsen}.}
  \bibinfo{year}{2016}\natexlab{}.
\newblock \showarticletitle{Third spaces in the age of IoT: a study on
  participatory design of complex systems}. In
  \bibinfo{booktitle}{\emph{Symposium on Challenges and experiences in
  designing for an ageing society, 12th International Conference on Designing
  Interactive Systems (COOP)}}.
\newblock


\bibitem[Thoppilan et~al\mbox{.}(2022)]%
        {thoppilan2022lamda}
\bibfield{author}{\bibinfo{person}{Romal Thoppilan}, \bibinfo{person}{Daniel
  De~Freitas}, \bibinfo{person}{Jamie Hall}, \bibinfo{person}{Noam Shazeer},
  \bibinfo{person}{Apoorv Kulshreshtha}, \bibinfo{person}{Heng-Tze Cheng},
  \bibinfo{person}{Alicia Jin}, \bibinfo{person}{Taylor Bos},
  \bibinfo{person}{Leslie Baker}, \bibinfo{person}{Yu Du}, {et~al\mbox{.}}}
  \bibinfo{year}{2022}\natexlab{}.
\newblock \showarticletitle{Lamda: Language models for dialog applications}.
\newblock \bibinfo{journal}{\emph{arXiv preprint arXiv:2201.08239}}
  (\bibinfo{year}{2022}).
\newblock


\bibitem[Ur et~al\mbox{.}(2014)]%
        {ur2014practical}
\bibfield{author}{\bibinfo{person}{Blase Ur}, \bibinfo{person}{Elyse McManus},
  \bibinfo{person}{Melwyn Pak Yong~Ho}, {and} \bibinfo{person}{Michael~L
  Littman}.} \bibinfo{year}{2014}\natexlab{}.
\newblock \showarticletitle{Practical trigger-action programming in the smart
  home}. In \bibinfo{booktitle}{\emph{Proceedings of the SIGCHI conference on
  human factors in computing systems}}. \bibinfo{pages}{803--812}.
\newblock


\bibitem[Ur et~al\mbox{.}(2016)]%
        {ur2016trigger}
\bibfield{author}{\bibinfo{person}{Blase Ur}, \bibinfo{person}{Melwyn Pak
  Yong~Ho}, \bibinfo{person}{Stephen Brawner}, \bibinfo{person}{Jiyun Lee},
  \bibinfo{person}{Sarah Mennicken}, \bibinfo{person}{Noah Picard},
  \bibinfo{person}{Diane Schulze}, {and} \bibinfo{person}{Michael~L Littman}.}
  \bibinfo{year}{2016}\natexlab{}.
\newblock \showarticletitle{Trigger-action programming in the wild: An analysis
  of 200,000 ifttt recipes}. In \bibinfo{booktitle}{\emph{Proceedings of the
  2016 CHI Conference on Human Factors in Computing Systems}}.
  \bibinfo{pages}{3227--3231}.
\newblock


\bibitem[Van~Brummelen et~al\mbox{.}(2020)]%
        {van2020convo}
\bibfield{author}{\bibinfo{person}{Jessica Van~Brummelen},
  \bibinfo{person}{Kevin Weng}, \bibinfo{person}{Phoebe Lin}, {and}
  \bibinfo{person}{Catherine Yeo}.} \bibinfo{year}{2020}\natexlab{}.
\newblock \showarticletitle{Convo: What does conversational programming need?}.
  In \bibinfo{booktitle}{\emph{2020 IEEE Symposium on Visual Languages and
  Human-Centric Computing (VL/HCC)}}. IEEE, \bibinfo{pages}{1--5}.
\newblock


\bibitem[Warnell et~al\mbox{.}(2018)]%
        {warnell2018deep}
\bibfield{author}{\bibinfo{person}{Garrett Warnell}, \bibinfo{person}{Nicholas
  Waytowich}, \bibinfo{person}{Vernon Lawhern}, {and} \bibinfo{person}{Peter
  Stone}.} \bibinfo{year}{2018}\natexlab{}.
\newblock \showarticletitle{Deep tamer: Interactive agent shaping in
  high-dimensional state spaces}. In \bibinfo{booktitle}{\emph{Proceedings of
  the AAAI conference on artificial intelligence}}, Vol.~\bibinfo{volume}{32}.
\newblock


\bibitem[Wei et~al\mbox{.}(2022)]%
        {wei2022chain}
\bibfield{author}{\bibinfo{person}{Jason Wei}, \bibinfo{person}{Xuezhi Wang},
  \bibinfo{person}{Dale Schuurmans}, \bibinfo{person}{Maarten Bosma},
  \bibinfo{person}{Ed Chi}, \bibinfo{person}{Quoc Le}, {and}
  \bibinfo{person}{Denny Zhou}.} \bibinfo{year}{2022}\natexlab{}.
\newblock \showarticletitle{Chain of thought prompting elicits reasoning in
  large language models}.
\newblock \bibinfo{journal}{\emph{arXiv preprint arXiv:2201.11903}}
  (\bibinfo{year}{2022}).
\newblock


\bibitem[Welleck et~al\mbox{.}(2019)]%
        {welleck2019neural}
\bibfield{author}{\bibinfo{person}{Sean Welleck}, \bibinfo{person}{Ilia
  Kulikov}, \bibinfo{person}{Stephen Roller}, \bibinfo{person}{Emily Dinan},
  \bibinfo{person}{Kyunghyun Cho}, {and} \bibinfo{person}{Jason Weston}.}
  \bibinfo{year}{2019}\natexlab{}.
\newblock \showarticletitle{Neural text generation with unlikelihood training}.
\newblock \bibinfo{journal}{\emph{arXiv preprint arXiv:1908.04319}}
  (\bibinfo{year}{2019}).
\newblock


\bibitem[Wu et~al\mbox{.}(2022)]%
        {wu2022ai}
\bibfield{author}{\bibinfo{person}{Tongshuang Wu}, \bibinfo{person}{Michael
  Terry}, {and} \bibinfo{person}{Carrie~Jun Cai}.}
  \bibinfo{year}{2022}\natexlab{}.
\newblock \showarticletitle{Ai chains: Transparent and controllable human-ai
  interaction by chaining large language model prompts}. In
  \bibinfo{booktitle}{\emph{Proceedings of the 2022 CHI Conference on Human
  Factors in Computing Systems}}. \bibinfo{pages}{1--22}.
\newblock


\bibitem[Yusuf et~al\mbox{.}(2022)]%
        {yusuf2022accurate}
\bibfield{author}{\bibinfo{person}{Imam Nur~Bani Yusuf},
  \bibinfo{person}{Lingxiao Jiang}, {and} \bibinfo{person}{David Lo}.}
  \bibinfo{year}{2022}\natexlab{}.
\newblock \showarticletitle{Accurate generation of trigger-action programs with
  domain-adapted sequence-to-sequence learning}. In
  \bibinfo{booktitle}{\emph{Proceedings of the 30th IEEE/ACM International
  Conference on Program Comprehension}}. \bibinfo{pages}{99--110}.
\newblock


\bibitem[Zeng(2019)]%
        {zeng2019learning}
\bibfield{author}{\bibinfo{person}{Andy Zeng}.}
  \bibinfo{year}{2019}\natexlab{}.
\newblock \emph{\bibinfo{title}{Learning visual affordances for robotic
  manipulation}}.
\newblock \bibinfo{thesistype}{Ph.\,D. Dissertation}.
  \bibinfo{school}{Princeton University}.
\newblock


\bibitem[Zeng et~al\mbox{.}(2022)]%
        {zeng2022socratic}
\bibfield{author}{\bibinfo{person}{Andy Zeng}, \bibinfo{person}{Adrian Wong},
  \bibinfo{person}{Stefan Welker}, \bibinfo{person}{Krzysztof Choromanski},
  \bibinfo{person}{Federico Tombari}, \bibinfo{person}{Aveek Purohit},
  \bibinfo{person}{Michael Ryoo}, \bibinfo{person}{Vikas Sindhwani},
  \bibinfo{person}{Johnny Lee}, \bibinfo{person}{Vincent Vanhoucke},
  {et~al\mbox{.}}} \bibinfo{year}{2022}\natexlab{}.
\newblock \showarticletitle{Socratic models: Composing zero-shot multimodal
  reasoning with language}.
\newblock \bibinfo{journal}{\emph{arXiv preprint arXiv:2204.00598}}
  (\bibinfo{year}{2022}).
\newblock


\bibitem[Zhang et~al\mbox{.}(2020)]%
        {zhang2020trace2tap}
\bibfield{author}{\bibinfo{person}{Lefan Zhang}, \bibinfo{person}{Weijia He},
  \bibinfo{person}{Olivia Morkved}, \bibinfo{person}{Valerie Zhao},
  \bibinfo{person}{Michael~L Littman}, \bibinfo{person}{Shan Lu}, {and}
  \bibinfo{person}{Blase Ur}.} \bibinfo{year}{2020}\natexlab{}.
\newblock \showarticletitle{Trace2tap: Synthesizing trigger-action programs
  from traces of behavior}.
\newblock \bibinfo{journal}{\emph{Proceedings of the ACM on Interactive,
  Mobile, Wearable and Ubiquitous Technologies}} \bibinfo{volume}{4},
  \bibinfo{number}{3} (\bibinfo{year}{2020}), \bibinfo{pages}{1--26}.
\newblock


\end{thebibliography}
\end{document}